%% file: 0-main-ndss-fall-2026.tex
\documentclass[conference]{IEEEtran}
\ifCLASSINFOpdf
\else
\fi

\usepackage{graphicx} 
\usepackage{array}    
\usepackage{comment}
\usepackage{makecell}
\usepackage{comment}
\usepackage{enumitem}
\usepackage{booktabs}
\usepackage{color}
\usepackage{xcolor}
\usepackage{soul}
\usepackage{url}
\usepackage{xspace}
\usepackage{listings}
\usepackage{tabularray}
\usepackage{tabularx}
\usepackage{xparse}
\usepackage{nicefrac}
\usepackage[available,functional,reproduced]{ndssbadges}

\usepackage{hyperref}
\usepackage{mycommands}
\definecolor{codegray}{rgb}{0.95,0.95,0.90}
\definecolor{commentgreen}{rgb}{0.0,0.5,0.0}

\usepackage[colorinlistoftodos,prependcaption,textsize=small]{todonotes}

\newcommand{\revnote}[2]{\textcolor{blue!70!black}{#1}\todo[color=blue!15]{#2}}

\newcommand{\tcnumber}[1]{\textcolor{gray}{TC-#1:}}

\lstdefinestyle{TC_commands}{
    language=Python,
    numbers=left,
    stepnumber=1,
    numbersep=5pt,
    basicstyle=\normalsize\ttfamily,
    numberstyle=\small\color{gray}\tcnumber,
    xleftmargin=23pt,
    breaklines=true,
    frame=none,
}

\lstdefinestyle{TERMINAL_OUTPUT}{
    numbers=none,
    language=bash,
    xleftmargin=0pt,
    breaklines=true,
    basicstyle=\scriptsize\ttfamily,
    frame=single,
    captionpos=b
}

\lstdefinestyle{python_code}{
    language=Python,
    numbers=none,
    xleftmargin=0pt,
    breaklines=true,
    basicstyle=\scriptsize\ttfamily,
    frame=lines,
    keepspaces,
}

\usepackage{pgfplotstable}

\usetikzlibrary{
  shapes.geometric,
  arrows,
  external,
  pgfplots.groupplots,
  matrix
}

\pgfplotsset{compat=1.9}

\usepackage{pifont}

\usepackage{mathtools,}

\usepackage[most]{tcolorbox}

\newtcolorbox{myBox}[3][]{
arc=5mm,
lower separated=false,
fonttitle=\bfseries,
colbacktitle=green!10,
coltitle=green!50!black,
enhanced,
attach boxed title to top left={xshift=0.5cm,
        yshift=-2mm},
colframe=green!50!black,
colback=green!10,
overlay={

at (frame.north east) {#3};},
title=#2,#1}

\usepackage{mycommands}
\usepackage{survey-results}

\begin{document}
%
\title{HoneySat: A Network-based \\ Satellite Honeypot Framework}

\author{
\IEEEauthorblockN{
Efrén López-Morales\IEEEauthorrefmark{3}\textsuperscript{*},
Ulysse Planta\IEEEauthorrefmark{2}\textsuperscript{*},
Gabriele Marra\IEEEauthorrefmark{2},
Carlos González\IEEEauthorrefmark{4}\IEEEauthorrefmark{8},
Jacob Hopkins\IEEEauthorrefmark{6},\\
Majid Garoosi\IEEEauthorrefmark{2},
Elías Obreque\IEEEauthorrefmark{5},
Carlos Rubio-Medrano\IEEEauthorrefmark{6},
Ali Abbasi\IEEEauthorrefmark{2}
}

\IEEEauthorblockA{
{\small\textsuperscript{*}Equal contribution, joint first authors}
}
\IEEEauthorblockA{
\IEEEauthorrefmark{3}New Mexico State University \quad
\IEEEauthorrefmark{2}CISPA Helmholtz Center for Information Security \\
\IEEEauthorrefmark{6}Texas A\&M University–Corpus Christi \quad
\IEEEauthorrefmark{4}Universidad de Santiago de Chile \\
\IEEEauthorrefmark{5}Universidad de Chile \quad
\IEEEauthorrefmark{8}German Aerospace Center (DLR), Braunschweig, Germany
}
}

%


\IEEEoverridecommandlockouts
\makeatletter\def\@IEEEpubidpullup{6.5\baselineskip}\makeatother
\IEEEpubid{\parbox{\columnwidth}{
		Network and Distributed System Security (NDSS) Symposium 2026\\
		23-27 February 2026, San Diego, CA, USA\\
		ISBN 979-8-9919276-8-0\\
		https://dx.doi.org/10.14722/ndss.2026.240537\\
		www.ndss-symposium.org
}
\hspace{\columnsep}\makebox[\columnwidth]{}}


\maketitle
\pagestyle{plain}       
\setcounter{page}{1}    

\input{abstract}


%
\IEEEpeerreviewmaketitle

\input{1-introduction}

\input{2-background}

\input{3-threat-model}

\input{4-system-design}

\input{5-implementation}

\input{6-evaluation}

\input{7-discussion}

\input{8-conclusion}

\input{ethics-and-open-science}



\bibliographystyle{IEEEtran}
\bibliography{bib}
%



\appendices

\input{9-appendix}
\newpage

\input{13-ae-appendix}

\end{document}

%% file: abstract.tex
\begin{abstract}
Satellites are the backbone of mission-critical services
that enable our modern society to function, for example, GPS. 
For years, satellites were assumed to be secure because of their indecipherable architectures and the reliance on security by obscurity.
However, technological advancements have made these assumptions obsolete, paving the way for 
potential attacks.
Unfortunately, there is no way to collect data on satellite adversarial techniques, hindering the generation of intelligence 
that leads to the development of countermeasures. 

In this paper, we present \honeyname, the first high-interaction satellite honeypot framework, capable of convincingly simulating a real-world CubeSat, a type of Small Satellite (SmallSat).  
To provide evidence of \honeyname's effectiveness, we surveyed SmallSat operators and deployed \honeyname over the Internet.

Our results show that 90\% of satellite operators agreed that \honeyname 
provides a realistic simulation. Additionally, \honeyname successfully deceived adversaries in the wild and collected 22 real-world adversarial interactions.
Finally, we performed a hardware-in-the-loop operation where \honeyname successfully communicated with an in-orbit, operational \smallsat mission. 

\end{abstract}

%% file: 1-introduction.tex
\section{Introduction}\label{sec:introduction}

Satellites are complex devices designed to withstand outer space conditions. They serve multiple purposes or types of \emph{missions} 
that include position and navigation, e.g., the Global Positioning System (GPS) constellation~\cite{space2023global}.
%
In addition, spacecraft can range from being as massive as thousands of kilograms, such as the International Space Station (ISS), to much smaller one-kilogram \emph{CubeSats}~\cite{uwe1_wuerzburg_2005}. 
As a result, the software and hardware components that make up a specific spacecraft vary greatly.
A cyberattack on a satellite or satellite constellation (group of satellites) could have disastrous consequences on a global scale, which is difficult to comprehend. Such an attack could lead to the cessation of air traffic and widespread communication blackouts~\cite{holmes2024growing}. It could also cause food shortages and the freezing of financial transactions~\cite{vancamp2022world}.
Furthermore, such an attack could exacerbate the Kessler Syndrome~\cite{drmola2018kessler}, a scenario in which collisions between satellites and debris in orbit create a cascade effect, generating even more debris,  
jeopardizing 
future satellite launches and operations. 

In this increasingly vulnerable environment, the probability of a successful satellite cyberattack continues to rise. This is driven by three key trends~\cite{holmes2024growing}:
%
first, 
satellite deployments have increased at an unprecedented pace. For instance, while an average of 82 launches took place between 2008 and 2017, as many as 197 launches occurred in 2023 alone, each typically carrying multiple satellites~\cite{semanik2023private}.
%
Second, ground station technology has become significantly more affordable (and sometimes open source), greatly lowering the communication barrier with satellites~\cite{california2024earth}. Thus, a broader range of malicious actors can now communicate with satellites. 
Third, satellite engineers and operators continue to rely on 
\emph{protocol obscurity} practices, such as hiding specialized knowledge about the transmission protocol implemented on 
satellites~\cite{willbold2023space}.

Although satellite security research has gained increased attention~\cite{willbold2023space,scharnowski2023case}, the relationship between outer space and cyberspace remains poorly understood~\cite{pavur2022building}. Although the volume of reported cyber incidents in the space sector has grown, these reports rarely provide sufficient detail~\cite{vancamp2022world}. As a result, the security community has limited visibility into adversarial activity aimed at space infrastructure.

A honeypot's ability to collect real-world cyberattack data makes it an ideal solution to this problem~\cite{holz2005detecting}. A honeypot is a decoy computer system intended to lure and entice malicious actors to interact with it~\cite{cohen2006use}; all the while, the honeypot logs all interactions. These logs can be analyzed to discover Tactics Techniques and Procedures (TTPs).

Since the release of the first honeypot, the Deception Toolkit, in 1997~\cite{cohen1998deception}, a wide range of honeypots with ever-increasing capabilities have been introduced. These honeypots are used by universities, companies, and nation-states worldwide~\cite{franceschi2023thousands,burgess2023clever,hilt2020caught,stingar2024clever,lopez2022honeyplc}. Honeypots are also used to deter malicious actors from attacking different types of systems, from industrial control systems~\cite{lopez2022honeyplc} to social media platforms~\cite{acharya2024conning}. However, as of the time of writing, \emph{there is no space-sector specific honeypot} in the literature.

In this paper, we design, implement, and evaluate the first satellite honeypot, \emph{\honeyname}, to attract and analyze adversaries who attack space infrastructures over the Internet, a commonly observed threat vector~\cite{boschetti2022space,wireless2024bisping,kavallieratos2023exploratory,willbold2024vsaster}. \honeyname is a modular, high-interaction honeypot framework that realistically simulates a complete satellite system (ground infrastructure and satellite). Specifically, \honeyname simulates \emph{Small Satellites} or \emph{\smallsat{}s}, which are spacecraft with a mass of less than 180 kilograms~\cite{nasa2025smallsats}. CubeSats, for example, are \smallsat{}s. Additionally, as part of \honeyname, we developed the \emph{\apiname}, a Python project to provide generic simulation functionality for users to populate satellite honeypots with believable data. 

We leveraged our framework to create honeypots of real-world CubeSat missions. Our results, backed by our survey of satellite operators, show that \honeyname's simulation is highly realistic. Our framework simulates an entire satellite mission, provides realistic telemetry, and supports real telecommands. 

Finally, we collaborated with an aerospace company to perform a hardware-in-the-loop experiment where \honeyname successfully communicated with an in-orbit, operational \smallsat.

In summary, this paper makes the following contributions:

\begin{itemize}

    \item We introduce \honeyname, a novel, high-interaction, extensible honeypot framework for small satellites (Sec.~\ref{sec:architecture}).
    
    \item We present the \apiname, a library that simulates the physical processes, sensors, and subsystems necessary for a realistic satellite honeypot (Sec.~\ref{sec:implementation}).
    
    \item We demonstrate \honeyname's high level of realism through a hands-on survey where 10 experienced satellite operators interacted with \honeyname in \emph{real time} (Sec.~\ref{subsec:survey}).

    

    \item We show \honeyname's extensibility features by integrating two real-world satellite flight software and integrating \honeyname into a hardware-in-the-loop operation allowing our honeypot to communicate with an in-orbit operational \smallsat (Sec.~\ref{subsec:case-study} and Sec.~\ref{subsec:hardware-integration}).

    
\end{itemize}

In the spirit of open and reproducible science, \honeyname's source code and experimental results are available online\footnote{https://doi.org/10.5281/zenodo.17871431},\footnote{https://github.com/HoneySat}.

%% file: 2-background.tex
\section{Background}\label{sec:background}

This section lays out key background concepts that are relevant to satellite honeypots.  
For honeypots: 
a description of the different existing 
types~(Sec.~\ref{subsec:honeypots}), as well as the current
state-of-the-art~(Sec.~\ref{subsec:existinghoneypots}). For satellites: their operation~(Sec.~\ref{subsec:satellite-context}), architecture~(Sec.~\ref{subsec:background:satellite-arch}), existing
protocol ecosystems~(Sec.~\ref{subsec:ecosystems}), and tactics, techniques, and procedures (TTPs) (Sec.~\ref{subsec:ttps}).

\subsection{Types of Honeypots}\label{subsec:honeypots}

Honeypots are categorized by interaction levels according to the interaction opportunities they provide. The two main types of honeypots are low-interaction and high-interaction.

\textbf{Low-Interaction Honeypots}. These honeypots offer minimal interaction, simulating real systems through scripts. They are easy to setup and maintain, and have a reduced risk of adversarial takeover. However, they provide limited interaction opportunities, which limits the data they provide. Some examples include Conpot~\cite{vestegaard2014conpot} and Honeyd~\cite{provos2007virtual}.

\textbf{High-Interaction Honeypots}. These honeypots offer extensive interaction opportunities via emulation or simulation~\cite{provos2007virtual}. Their main advantage is providing adversaries with almost limitless interactions.
However, they pose a high takeover risk~\cite{marra2024feasibility}. Examples include Cowrie~\cite{oosterhof2024cowrie} and HoneyPLC~\cite{lopez2022honeyplc}.

\subsection{Honeypot's State of the Art}\label{subsec:existinghoneypots}

The literature on honeypots includes hundreds of implementations that simulate a diverse set of systems~\cite{ilg2023survey,nazario2024awesome}. From implementations that simulate a host's TCP/IP stack, such as Honeyd~\cite{provos2007virtual}, to modern approaches that integrate social media applications, such as HoneyTweet~\cite{acharya2024conning}. In the absence of a satellite honeypot, we examine the approaches most related to satellites: Industrial Control Systems (ICS) and Unmanned Aerial Vehicles (UAV), summarized in Table~\ref{table:literature-comparison}.

Satellite systems like ICS must be aware of some physical process, e.g., the sun's position, via \emph{sensors} to acquire data about the physical world. ICS honeypots, such as HoneyICS~\cite{lucchese2023honeyics}, simulate these physical processes. In addition to ICS, UAV honeypots such as HoneyDrone~\cite{daubert2018honeydrone} achieve this via simulations, and recreate attack scenarios.

\input{tables/comparison-existing-honeypots}

\subsection{Anatomy of a Satellite Mission}\label{subsec:satellite-context}

\begin{figure}[t]
    \centering
    \includegraphics[width=1.0\linewidth]{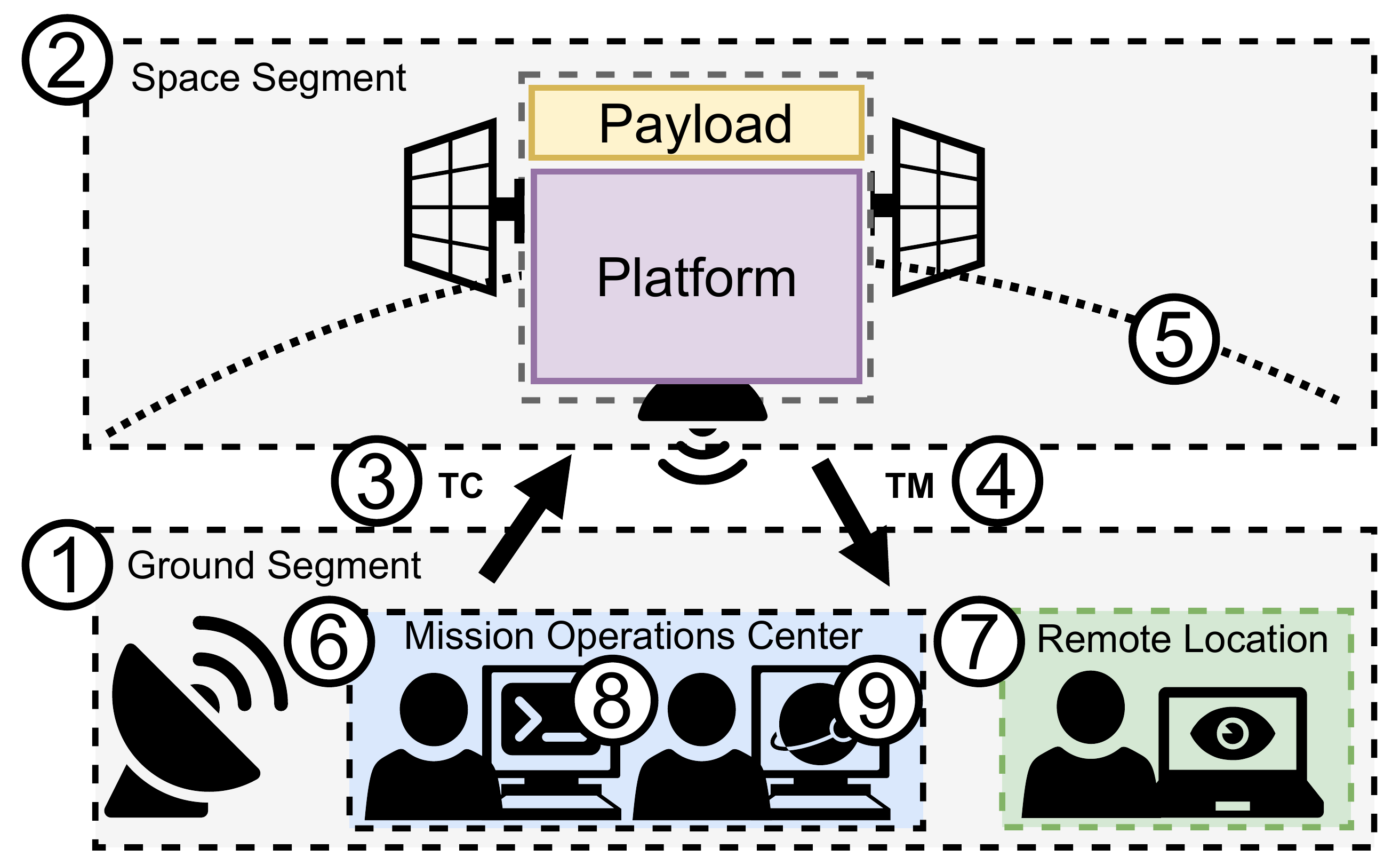}
    \caption{
    The 
    components commonly found in a satellite mission.}
    \label{fig:satellite-arch}
\end{figure}

We now describe the components of a satellite mission. Due to satellite missions' complexity and diversity, we explain each component and match it with one of the numbers in Fig.~\ref{fig:satellite-arch}. Every satellite mission includes the \emph{ground segment} from which satellite operators control the satellite and the \emph{space segment} which includes the satellite itself.

\textbf{Ground Segment \circled{1}}. The Ground Segment (GS) covers the terrestrial infrastructure required for a successful satellite operation. It consists of a ground station, responsible for exchanging data with the spacecraft, the computational and network infrastructure required for communication, and the systems to operate the satellites, e.g., servers~\cite{willbold2023space}. The GS includes the Ground Segment Software (GSS) that helps operators schedule and send commands and visualize data.

\textbf{Space Segment \circled{2}}. The space segment comprises a satellite or a constellation of satellites. A satellite is launched into orbit and then establishes communications with the ground segment.
During regular operations, satellites may communicate through one or multiple ground stations~\cite{willbold2023space}.

\textbf{Telecommands (TC) \circled{3} and Telemetry (TM) \circled{4}}. The basic data flow between the space and ground segments are TC and TM~\cite{esa2013telemetry}. TM is the data the satellite sends to the ground station which may contain the satellite's status or payload data~\cite{esa2013telemetry}. TCs are used to operate the satellite and are transmitted and encapsulated in a space protocol (see Sec.~\ref{subsec:ecosystems}). The design and implementation of TCs varies depending on the satellite mission. From a security perspective, TCs are particularly important as an attacker that can send valid TCs to a satellite can fully take over the mission~\cite{willbold2023space}.

\textbf{Orbital Pass \circled{5}}. 
A satellite and its ground station can communicate \emph{only} during an orbital pass. An orbital pass, or simply a \emph{pass}, is when the satellite rises above a ground station's horizon and becomes available for communication. A pass's duration and timing depend on the satellite's orbit characteristics and any obstructing objects, e.g., mountains~\cite{wood2006introduction}. Passes can be predicted using two-line element (TLE) data~\cite{vallado2012two}.

\textbf{Satellite Mission Operations \circled{6}}. Satellite mission operations vary widely depending on the owning organization's budget, and technology~\cite{maya1_inquirer_2018}. Nevertheless, they share some commonalities, which we now describe.

A mission's operation involves a team of operators that use ground segment software (GSS) to operate a satellite(s) and ensure the mission's success~\cite{nasa2024flight}.
Operations are carried out in a Mission Operations Center (MOC), where operators sit at their workstations to manage TCs sent to the satellite(s). 
Operators may also remotely operate satellites \circled{7} by connecting to the ground segment using tools such as VNC (Virtual Network Computing)~\cite{california2024earth}.

Satellite operations include two main activities: satellite tracking and TC generation and scheduling. 
Satellite tracking calculates the satellite's position in orbit and controls the ground station's tracking antenna to establish communication between the ground and space segments.
TC generation and scheduling crafts commands to be sent to the satellite to perform different functions, e.g., download payload data.

\textbf{Ground (Segment) Software (GSS)}. GSS allows operators to carry out the mission's routine operations.
GSS are very diverse. Some satellite missions develop their own GSS while others use open-source~\cite{suchai-flight-software} or proprietary GSS~\cite{gomspace2021nanocom}. There are two main types of GSS: Mission Control Software (MCS) and Ground Station Control Software (GSCS).

Mission control software \circled{8} manages TCs and scripts to be sent to the satellite and display TM. For example, ESA's SCOS-2000 is an MCS that provides generic functionality that can be customized for a specific mission operation~\cite{merri2007cutting}. Some missions develop their own MCS. 
For example, the 
SUCHAI mission developed their own MCS application~\cite{suchai-flight-software}.

Ground station control software (GSCS) \circled{9} helps satellite operators track and visualize the satellite's orbit and provide information about each satellite's pass. For example, Gpredict is a popular open-source GSCS that performs real-time satellite tracking and orbit prediction~\cite{csete2023gpredict}.

\subsection{Satellite Architecture}\label{subsec:background:satellite-arch}

Satellite architectures are varied and complex~\cite{ivancic2003architecture}; however, here we describe the most common terminology depicted in Fig.~\ref{fig:satellite-arch}'s space segment \circled{2}.

When referring to satellite architecture there is a distinction between the \emph{platform} that facilitates the successful operation of the satellite's activities and the \emph{payload}. The platform underpins the payload that fulfills the mission's purpose. Payloads differ depending on the satellite's mission and can range from measurement instruments to communications systems.

\textbf{Platform}: The platform is composed of custom-designed or off-the-shelf subsystems necessary for critical satellite operations. These include the Attitude Determination and Control System (ADCS) to maintain the satellite's orientation (i.e. attitude); the Electrical Power Subsystem (EPS) for managing power generation and distribution; the Communication Subsystem (COMM) and the Command and Data Handling (C\&DH) subsystems to facilitate communications for receiving TC and sending TM.

These subsystems are controlled via TCs sent to the satellite.
TCs may be interpreted by a central C\&DH System or merely forwarded to the recipient subsystem~\cite{yost2023nasa}.
This managerial duty is handled by the Flight Software (FS) running on an embedded system, for example, NASA's Core Flight System (cFS)~\cite{mccomas2016core}. Attackers aim to gain the ability to send TCs to the flight software as this may grant control over all subsystems.

\textbf{Payload}. The payload is the equipment that a satellite employs to fulfill its mission. Due to satellites' varied missions, payloads are heavily customized~\cite{esa2018payload}. For example, if a satellite's mission is remote sensing, its payload may include an infrared camera~\cite{nasa2019remote}.
\subsection{Small Satellite Protocol Ecosystems}\label{subsec:ecosystems}\label{subsec:comm-protocols}

\smallsat missions can often be categorized by the adoption of protocols and their corresponding philosophies. Currently, there are two major protocol ecosystems which \smallsat missions can adopt: CSP and CCSDS.

\textbf{Cubesat Space Protocol.}
The Cubesat Space Protocol (CSP) family of protocols is a one-stop solution for SmallSat missions~\cite{LibcspLibcsp2024}. CSP is implemented as an open-source C library called \emph{libCSP}~\cite{LibcspLibcsp2024} and follows the TCP/IP model, including transport and routing protocols and multiple layer 2 interfaces such as I\textsuperscript{2}C (Inter-Integrated Circuit), and ZeroMQ (ZMQ) for transmission on TCP/IP networks~\cite{GetStarted}.

In the context of a satellite mission, CSP connects the ground segment and satellite subsystems as part of a CSP network where each subsystem is identified as a node. Sending a TC is as simple as sending a CSP packet with the address of the corresponding subsystem node. 



\textbf{CCSDS Space Communication Protocols.} The Consultative Committee for Space Data Systems (CCSDS) Space Communication is a set of standardized protocols used for different purposes in space communications~\cite{pusecosysytem}. 
A relevant CCSDS protocol for \smallsat TM and TC is the Space Packet protocol. This protocol is used in combination with the ECSS Packet Utilization Standard (PUS)~\cite{kaufeler1994esa} to define how TCs and TM are encoded and transported. PUS defines services (and thus sets of TC/TM) for functionality that satellite missions require,
including large data transfer or event reporting~\cite{pusecosysytem}. 

\subsection{Space Systems' Tactics, Techniques and Procedures (TTPs)}\label{subsec:ttps}

Tactics, Techniques, and Procedures (TTPs) describe a malicious actor's behavior in a structured scheme to understand how they might execute an attack~\cite{raza2024what,nist2024tactics}. The SPACE-SHIELD matrix~\cite{esa2024space-shield} is a framework used to standardize space systems' TTPs, for example, \emph{ground segment compromise}.

%% file: tables/comparison-existing-honeypots.tex
\definecolor{green-dark}{RGB}{0,100,0}

\begin{table}[t]
\centering
\caption{Comparison of 
Existing 
Honeypots and \honeyname.}
\begin{center}
{\footnotesize
Keys:
\fullmark\xspace =\xspace Supported;
\emptymark\xspace=\xspace Not Supported.
}
\end{center}  

\label{table:literature-comparison}
\setlength{\tabcolsep}{1.5pt} 
\renewcommand{\arraystretch}{0.7} 
\begin{tabular}{lcccc}

\toprule
\textbf{\begin{tabular}[c]{@{}c@{}}Honeypot /\\ Feature\end{tabular}}                         
& \textbf{\begin{tabular}[c]{@{}c@{}}Interaction\\Level \end{tabular}}
& \textbf{\begin{tabular}[c]{@{}c@{}}Included \\ Protocols\end{tabular}} 
& \textbf{\begin{tabular}[c]{@{}c@{}}Physics Sims\end{tabular}}      
& \textbf{\begin{tabular}[c]{@{}c@{}}Extensibility\end{tabular}} 
\\ 
\bottomrule
\begin{tabular}[c]{@{}c@{}}
Conpot  \cite{vestegaard2014conpot}
\end{tabular}                                                                   
& Low   
& 9
& 0
& \fullmark

\\
\begin{tabular}[c]{@{}c@{}}
HoneyPLC\cite{lopez2022honeyplc}
\end{tabular}     
& High                                                                      
& 3
& 0    
& \fullmark

\\
\begin{tabular}[c]{@{}c@{}}
ICSPot  \cite{conti2022icspot}
\end{tabular}
& High              
& 4                               
& 1            
& \emptymark
\\
\begin{tabular}[c]{@{}c@{}}
HoneyICS\cite{lucchese2023honeyics}
\end{tabular}
                                                              
& High                                                             
& 2                     
& 1 
& \fullmark

\\
\toprule
\begin{tabular}[c]{@{}c@{}}HoneyDrone \cite{daubert2018honeydrone}\end{tabular}       
& Medium		
& 4 						       							  
& 1     	
& \emptymark

\\
\toprule
\begin{tabular}[c]{@{}c@{}} \textcolor{green-dark}{\textbf{\honeyname}} \end{tabular}           		
&\begin{tabular}[c]{@{}c@{}} \textcolor{green-dark}{High} \end{tabular}
&\begin{tabular}[c]{@{}c@{}}
\textcolor{green-dark}{4}

\end{tabular}
&\begin{tabular}[c]{@{}c@{}} \textcolor{green-dark}{6} \end{tabular}
&\begin{tabular}[c]{@{}c@{}} \fullmark \end{tabular}
\\
\toprule

\begin{tabular}[c]{@{}c@{}} Addressed \\in Section \\  \end{tabular}

&\begin{tabular}[c]{@{}c@{}}
\ref{subsec:space-design}, \\ \ref{subsec:ground-design}
\end{tabular}

&\begin{tabular}[c]{@{}c@{}} \ref{subsec:ground-design}, \\ \ref{subsec:ground:implementation}
\end{tabular}

&\begin{tabular}[c]{@{}c@{}}
\ref{subsec:space-design}, \\ \ref{subsec:satellitesimimplementation}
\end{tabular}

&\begin{tabular}[c]{@{}c@{}} \ref{subsec:case-study}\\\ref{subsec:hardware-integration}
\end{tabular}


\\
\bottomrule
\end{tabular}

\end{table}

%% file: 3-threat-model.tex
\section{Threat Model}\label{sec:threat-model}

Following Fig.~\ref{fig:honeysat-arch}, 
we assume that an adversary willing to compromise a satellite can only interact with the space segment simulation by sending TCs from the ground segment first.
To gain initial access to the ground segment, an adversary needs to connect via one of the exposed network protocols discussed in Sec.~\ref{subsec:comm-protocols}, which correspond to the operational protocols used in real satellite missions. From there, an adversary may try to launch different commands to take full control and/or compromise the services offered by the satellite's mission as depicted in Fig.~\ref{fig:theoryoperation}. 
Finally, in this paper, the modeling of physical radio communication between the space and ground segments, which is commonly used in practice, is considered out of scope and left for future work. The threat model is further referenced in Sec.~\ref{subsec:theoryoperation}.

%% file: 4-system-design.tex
\section{\honeyname Framework Design}\label{sec:architecture}
In this section, we explain the objectives we aim to achieve~(Sec.~\ref{subsec:design:objectives}), the design principles 
we follow to meet such objectives~(Sec.~\ref{subsec:design:principles}), and the overall design of 
our 
Space
~(Sec.~\ref{subsec:space-design}), and Ground (Sec.~\ref{subsec:ground-design}) segments.

\subsection{\honeyname's Design Objectives}\label{subsec:design:objectives}

Our design aims to achieve the following objectives:

\begin{objectives}

\item \label{objective:rich-interactions}
\textbf{Capability to Capture Rich Interaction Data.} As explained in Sec.~\ref{sec:introduction}, the number one objective of any honeypot is to capture interaction data from which we derive knowledge on adversaries' TTPs. As such, \honeyname's first objective is to capture rich interaction data.

\item \label{objective:deception}
\textbf{Provide Deception.} As we discussed in Sec.~\ref{sec:introduction}, honeypots' nature must remain \emph{covert} to entice adversaries to interact with it. As such, our second objective is for \honeyname's nature to remain hidden from adversaries.

\item \label{objective:extensibility}
\textbf{Provide Extensibility and Customizability.}
A framework's main purpose is to provide generic functionality that can be customized to meet the user's needs. In \honeyname's case, we must be able to support multiple \smallsat{}s. For example, a particular \smallsat{} may use CSP or CCSDS ecosystems. As such, \honeyname's third objective is extensibility and customizability.

\end{objectives}

\begin{figure}[t]
    \centering
    \includegraphics[width=1.0\linewidth]{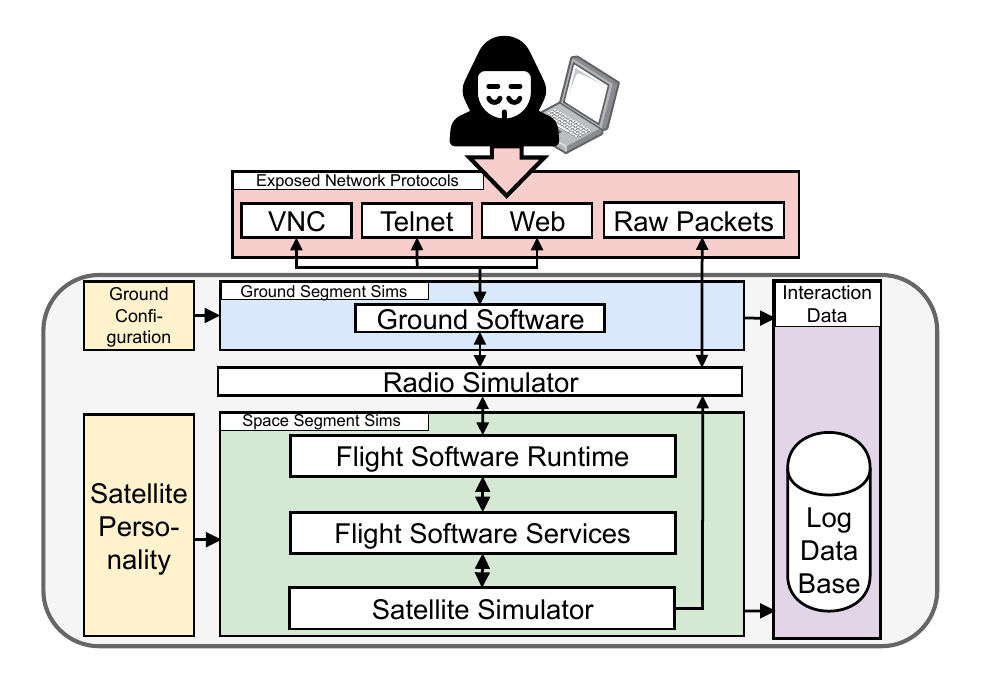}
    \caption{\honeyname framework high-level architecture design.
    }
    \label{fig:honeysat-arch}
\end{figure}

\subsection{\honeyname's Design Principles}\label{subsec:design:principles}

To meet our objectives, we selected the following principles:  

\begin{principles}

\item \label{principle:high}
\textbf{High-Interaction Simulation.} As we described in Sec.~\ref{subsec:honeypots}, high-interaction honeypots give adversaries the same or almost identical interaction opportunities as a real satellite. For these reasons, we selected high-interaction simulation as our first design principle.
This design principle is based on design objective~\ref{objective:rich-interactions}.

\item \label{principle:realistic}

\textbf{Realistic Simulation.} As discussed in Sec.~\ref{subsec:existinghoneypots}, satellites track their orbit location, among others. These details must be simulated; otherwise, they may alert adversaries that they are interacting with a fake system. This design principle is based on design objective~\ref{objective:deception}.

\item \label{principle:modular}
\textbf{Modularity.} As we explained in Sec.~\ref{sec:background}, satellites are complex and diverse systems. To tackle this problem, we designed \honeyname to be modular. This design principle is based on 
our
objective~\ref{objective:extensibility}.

\end{principles}

We now describe \honeyname's architecture design and how this architecture integrates the \emph{design principles} described above. At the highest level, our framework's architecture, depicted in Fig.~\ref{fig:honeysat-arch}, consists of two sets of simulations, the \emph{space segment simulations} and the \emph{ground segment simulations}. Table~\ref{table:real-components-comparison} illustrates how \honeyname's simulation components match the components of a real satellite mission.

\subsection{Ground Segment Design}\label{subsec:ground-design}

Following 
Sec.~\ref{subsec:satellite-context}, the purpose of 
the \honeyname's ground segment (depicted in Fig.~\ref{sec:architecture}) is to simulate the ground segment assets, e.g., the 
ground segment software. To accomplish this, our design includes the following components:
1)~the \emph{Exposed Network Protocols}, 
2)~the \emph{Ground Segment Software}, 
3)~the \emph{\radioname}, 
4)~the \emph{Ground Personality}, and 
5)~the \emph{Logging Repository}.

\textbf{1) Exposed Network Protocols.} 
To provide adversaries with feasible access to our honeypot, \honeyname exposes multiple means of interaction over a network such as the Internet.
We designed \honeyname to support four different interaction methods using different protocols, namely VNC, Telnet, Web, and Access to the ground station via raw packet transmitting capability; following the design principle~\ref{principle:modular}. We selected these protocols based on data obtained from our satellite operator survey, discussed in Sec.~\ref{subsec:survey}, which revealed that satellite operators do use remote access tools, such as web interfaces and screen sharing, to operate satellites.


\textbf{2) Ground Software.} As discussed in Sec.~\ref{subsec:satellite-context}, the ground software includes mission control software (MCS) and ground station control software (GSCS). We leverage existing ground software, e.g., Gpredict, allowing \honeyname to provide a high-interaction simulation following design principle \ref{principle:high}.

\textbf{3) Radio Simulator.} As discussed in Sec.~\ref{subsec:satellite-context}, communication between the satellite and the ground station is possible only during a pass. 
The purpose of the \radioname is to mimic real orbital passes by enabling and disabling communication between \honeyname's ground and space segment simulations at the appropriate times. The \radioname design follows the design principles~\ref{principle:realistic} and~\ref{principle:modular}.

\textbf{4) Ground Configuration.} The ground configuration is a series of settings for the ground segment-specific configurations, e.g., the satellite mission logo in the Web Interface.

\textbf{5) Logging Repository.} This repository records data such as the TM/TC traffic to and from the ground station software and the logging attempts received in the web interface. 
We designed the ground segment simulation logs to be categorized and timestamped. The logging repository design follows design principle \ref{principle:modular}.

\subsection{Space Segment Design}\label{subsec:space-design}

The purpose of the  
\honeyname's space segment is to mimic the spacecraft, as discussed in Sec.~\ref{subsec:satellite-context}. To accomplish this, our design includes the following components: 
1)~the satellite's \emph{Flight Software Runtime}, 
2)~the \emph{\apiname}, 
3)~the \emph{Flight Software Services}, 
4)~the \emph{Satellite Personality}, and 
5)~the \emph{Logging Repository}. \\

\textbf{1) Flight Software Runtime.
} As discussed in Sec.~\ref{subsec:background:satellite-arch}, the satellite flight software manages all critical functions required for the mission operation, such as interacting with hardware peripherals, processing TCs, and sending TM.
For this to work we need an environment where services that handle TC intended to run on flight software can be run.
We reuse existing compatibility or testing wrappers to run the relevant parts of flight software for \honeyname to provide nearly identical interactions to an adversary. This produces a high-interaction honeypot simulation that follows design principle \ref{principle:high}.


\begin{figure}[t]
    \centering
    \includegraphics[width=1.0\linewidth]{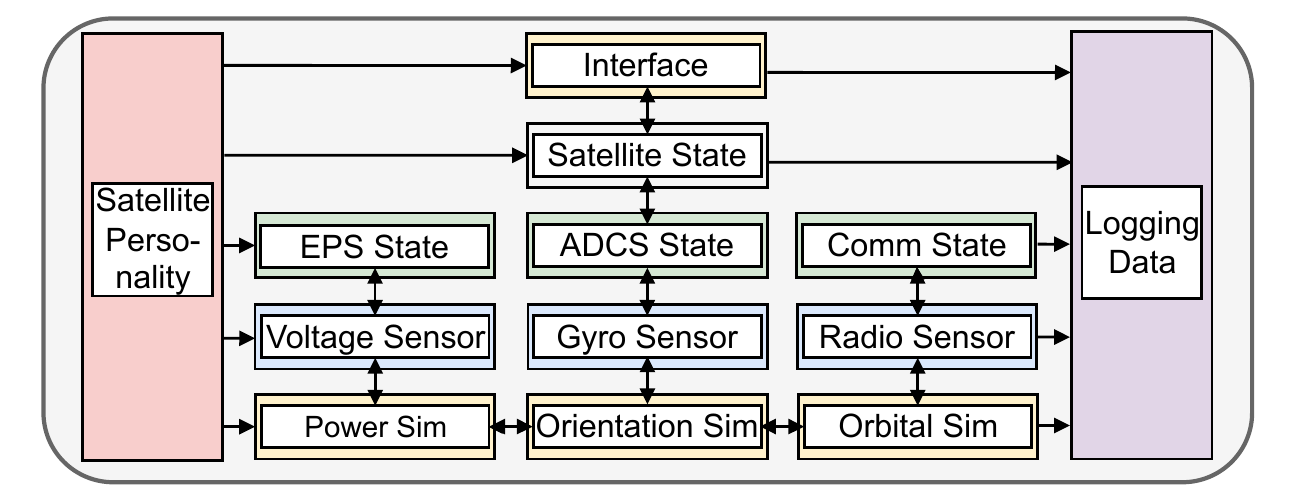}
    \caption{High-level architecture of the \apiname. Due to space limitations, we do not show all the available sensors and simulations. 
    }
    \label{fig:honeysat-api-arch}
\end{figure}

\textbf{2) \apiname.} One of the biggest challenges when designing \honeyname was simulating all the physical processes, e.g., attitude, that satellites need to know. The \apiname solves this problem by simulating 
all the necessary satellite subsystems, e.g., the Electrical Power System, and the sensors, e.g., a GPS receiver. 

As shown in Fig.~\ref{fig:honeysat-api-arch}, the \apiname includes five module types: \emph{simulation modules} (orange), \emph{sensor modules} (blue), \emph{subsystem state modules} (green), \emph{satellite state modules} (gray) and \emph{interface modules} (yellow).

\emph{Simulation Modules}. 
These modules are abstractions of real-world processes required for a realistic satellite simulation (\ref{principle:realistic}), e.g., the orbital simulation which uses orbital mechanics to calculate data like the satellite's orbital position.

These simulation modules can communicate with each other to facilitate proper functionality. For example, the power simulation queries the orbital simulation for data to determine if the satellite is positioned properly to draw power from its solar panels. The \apiname includes six simulations, the orbital, rotation, power system, thermal, magnetic, and payload simulations.


\begin{itemize}[leftmargin=*]
    \item \emph{Orbital Simulation.} The orbital simulation calculates the satellite's orbit~\cite{vallado2012two}. It uses the TLE data configured in the satellite personality and it generates multiple values such as, latitude, and longitude of the satellite. The payload, EPS, and magnetic simulations rely on data from this simulation.

    \item \emph{Rotation Simulation.} This simulation calculates the satellite's attitude changes to provide its orientation by using a reference frame fixed to the satellite body and a non-rotational reference frame. The relation between the two reference frames is calculated using the conservation of angular momentum and the rigid-body Euler equation~\cite{euler}. 
    This simulation's data is used by payloads like the Red, Green, Blue (RGB) camera to point towards Earth's surface.

    \item \emph{Power System Simulation.}
    This simulation manages the satellite's power collection, consumption, and distribution.
    It tracks the battery capacity, the power draw and simulates battery charging whenever the orbital simulation tells it that the satellite is exposed to sunlight.

    \item \emph{Thermal Simulation.} This simulation calculates the satellite's temperature based on the total thermal energy and a user-defined specific heat capacity and uses a radiation loss formula~\cite{stefan-boltzmann} to calculate the thermal radiation emission. 
    

    \item \emph{Magnetic Simulation.} This simulation tracks and analyzes the interactions between the Earth's magnetic field and the satellite's own magnetic environment. It communicates with the orbital and rotation simulations to determine the satellite’s position and orientation to calculate the Earth's magnetic field components.

    \item \emph{RGB Camera Simulation.} This simulation replicates the functionality of an RGB camera pointed at Earth.
    It uses Earth observation satellite imagery from the U.S. Geological Survey (USGS)~\cite{geological2022earthexplorer}. If a capture image command is issued, it will select and return an image.
    This simulation communicates with the orbital simulation to determine if satellite is in the presence of the sun.

\end{itemize}


\emph{Sensor Modules.}  
These modules are abstractions of the hardware sensors used by a satellite, i.e., temperature sensor. 
The sensor modules do not perform any computations, instead they collect data directly from the simulation modules following \ref{principle:realistic}. For example, the voltage sensor will query the power simulation to collect data about the current state of the battery.

\emph{Subsystem State Modules.}
These modules are abstractions of the satellite subsystems discussed in Sec.~\ref{subsec:background:satellite-arch}. A given subsystem state is made up of one or more sensors. For example, in Fig.~\ref{fig:honeysat-api-arch} the \emph{Payload State} (light green) 
includes the \emph{Camera Sensor} (light blue). 
In this way, subsystem states serve two purposes. They can \emph{get} data from their sensors or can \emph{set} a specific configuration on their sensors.

\emph{Satellite State Module.} 
This module 
is an abstraction of an entire satellite. As depicted in Fig.~\ref{fig:honeysat-api-arch} the \emph{Satellite State} (light gray) is made up of multiple subsystem states. Satellite State module routes messages between the \apiname and the modified flight software. For example, if the modified flight software sends a message to the \apiname requesting to provide the present voltage in the EPS, the satellite state will route that message to the EPS State.

\emph{Interface Module.}
This module is an abstraction of the communication protocol between the modified flight software and the \apiname. As depicted in Fig.~\ref{fig:honeysat-api-arch}, the \emph{Interface} (light yellow) is the module that connects the flight software services with the \apiname simulations.

This protocol must implement two basic message types, \emph{requests} and \emph{replies}. However, to meet \ref{objective:extensibility}, the underlying implementation of these messages is left open for the user to decide based on their requirements.

\textbf{3) Flight Software Services.} The flight software services that process TC and provide TM act as the bridge between the \apiname and the rest of the honeypot. In case a service requires some data from a subsystem of the satellite or a command is sent to a subsystem, it is passed to the satellite simulator instead. The \emph{satellite state module} routes the service's messages to and from the \apiname. It uses a list of message IDs that can be customized based on the protocol ecosystem and software architecture complying with our design principles \ref{principle:high} and \ref{principle:modular}.

\textbf{4) Satellite Personality.} The satellite personality module is designed to provide a central configuration location for the space segment.
It includes FS and \apiname configuration values such as the satellite's battery capacity. The satellite personality makes our framework easily customizable; thus following our design principle \ref{principle:modular}.

\textbf{5) Logging Repository.} We designed each space segment component to provide detailed logs. As shown in Fig.~\ref{fig:honeysat-api-arch}, all \apiname modules can send their own logs.

\subsection{Theory of Operation}\label{subsec:theoryoperation}

\begin{figure}[t]
    \centering
    \includegraphics[width=1.0\linewidth]{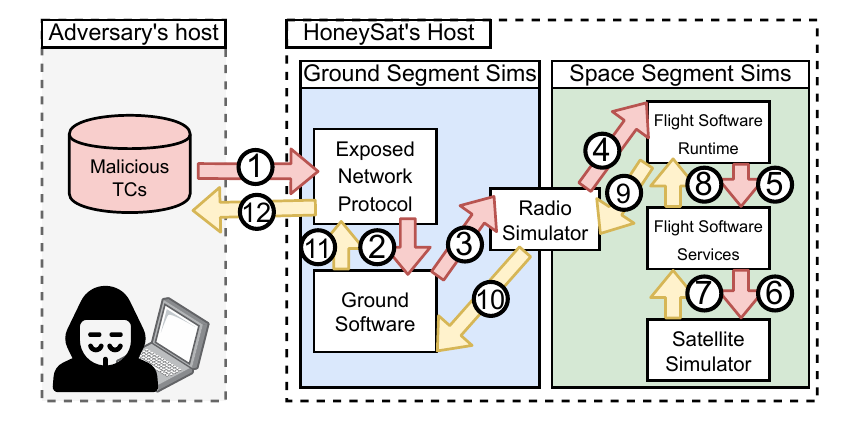}
    \caption{\honeyname's Theory of Operation. 
    }
    \label{fig:theoryoperation}

\end{figure}

In this section, we provide a brief overview of how the designed framework components shown in Figure~\ref{fig:honeysat-arch} interact by following the numbers in Fig.~\ref{fig:theoryoperation}.

An adversary gets initial access via one of the \emph{Exposed Network Protocols} \circled{1}. On interacting with the \emph{Simulated Ground Segment}, an attacker is presented with access to the \emph{Ground Software} \circled{2}. The configuration of this ground software is defined by the \emph{Ground Configuration}, which allows the ground software to look like one of many different missions from its protocol ecosystem to the attacker. The attacker can then interact with the \emph{Ground Software}, while their actions are reported to the \emph{Logging Repository}. They might want to try to gain more privileges on the ground segment or try to send TC to the space segment. Instead of deploying a ground station with an RF transmitter and associated hardware, we employ the \emph{Radio Simulator} \circled{3} to handle all simulated radio frequency communications.

When an attacker uses the ground segment to send a valid TC or raw packets, the \emph{Radio Simulator} checks whether the satellite configured in the \emph{Satellite Personality} is currently passing over the designated ground station location. If so, the \emph{Radio Simulator} forwards the TC to the \emph{Flight Software Runtime} \circled{4}. When the \emph{Flight Software Runtime} receives a command, it will run the corresponding \emph{Flight Software Service} to compute a response \circled{5}. In case it requires either changes to an on-board system’s state or a sensor value to execute the TC, it will invoke the \emph{Satellite Simulator} \circled{6}. The \emph{Satellite Personality} contains the configuration used by the \emph{Satellite Simulator}. Once a TC is processed, the resulting TM is routed back through the \emph{Radio Simulator} to either the attacker or the ground software used by the attacker \circled{7}-\circled{12}, and is also recorded in our \emph{Logging Repository}.


%% file: 5-implementation.tex
\section{\honeyname's  Implementation}\label{sec:implementation}

Having laid out \honeyname's design we now explain how we implemented the \apiname (Sec.~\ref{subsec:satellitesimimplementation}), the generic CSP mission honeypot (Sec.~\ref{subsec:cspgeneric}), the ground segment (Sec.~\ref{subsec:ground:implementation}), and the space segment (Sec.~\ref{subsec:space:implementation}).


\subsection{\apiname Implementation}\label{subsec:satellitesimimplementation}

We implemented the \apiname as a Python object-oriented programming application. 
We implemented \sensors, \subsystems, 1 satellite state, and 1 interface, based on our \simulations. 



\subsection{Generic CSP Mission Honeypot}\label{subsec:cspgeneric}

We implemented \honeyname to support the CSP ecosystem providing a generic CSP-based honeypot that can simulate any specific CSP-based mission.

\subsubsection{Ground Segment Implementation}\label{subsec:ground:implementation}

We now describe how we implemented the five ground segment simulations designed in Sec.~\ref{subsec:ground-design}.
1)~the \emph{Exposed Entry Points}, 
2)~the \emph{Ground Software}, 
3)~the \emph{Radio Proxy}, 
4)~the \emph{Ground Configuration}, and 
5)~\emph{Logging Repository}.

\textbf{1) Exposed Network Protocols.} We implemented a VNC server, a Telnet server, a Web server, and ZeroMQ-based access to the CSP Network. We implemented the VNC server using TigerVNC~\cite{ossman2024tigervnc} and used PyZMQ~\cite{granger2024pyzmq} for the raw packet access. The Telnet server was implemented using Python to expose the command-line interface of the Ground Segment Software. The web interface allows for the customization of text presented to the attacker through configuration. 

\begin{figure}[t]
    \centering
    \includegraphics[width=1.0\linewidth]{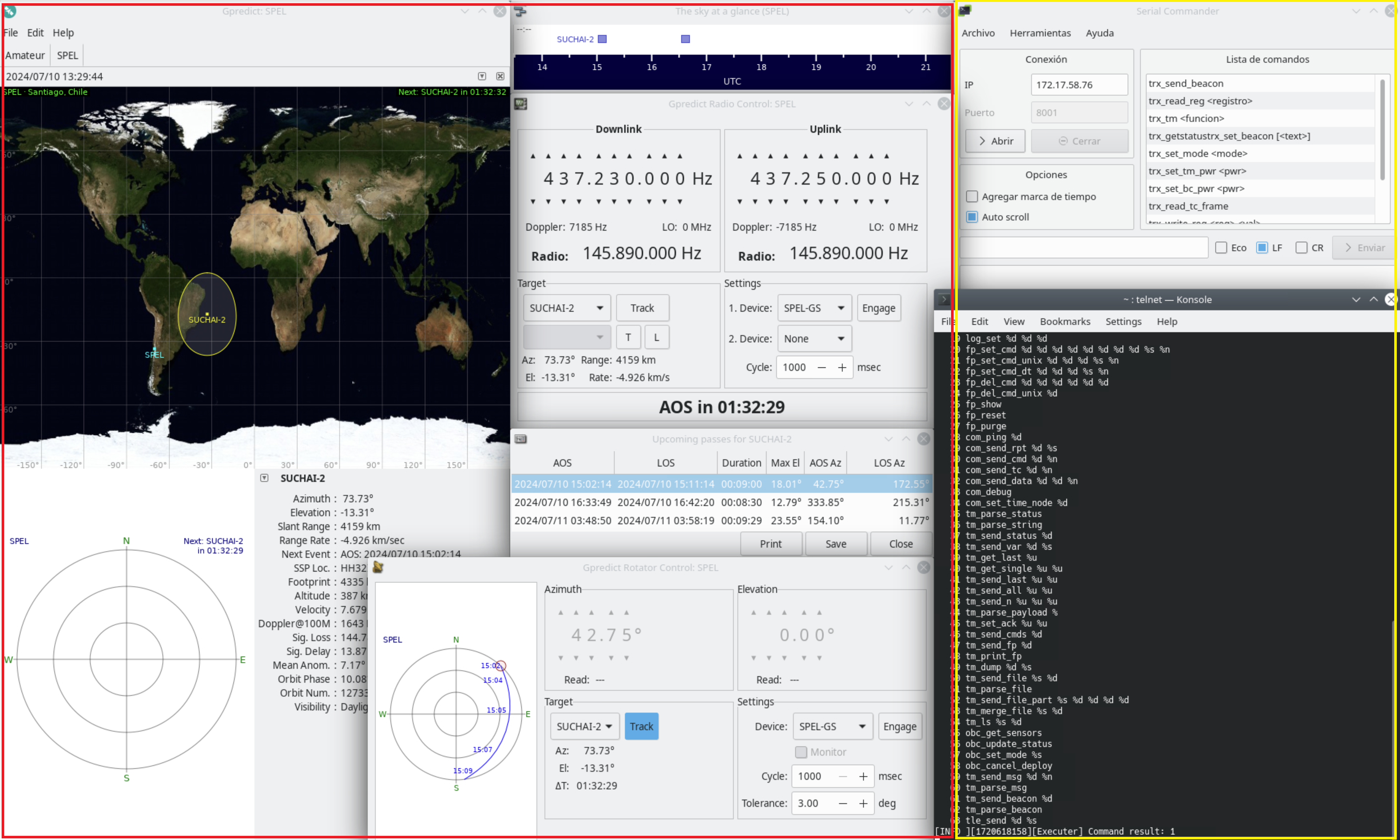}
    \caption{The view an attacker would see upon connection to the VNC server. Ground Station Control Software (Red) Ecosystem/Mission Specific MCS (Yellow).}
    \label{fig:suchai-vnc-honeypot} 
\end{figure}


\textbf{2) Ground Software.} To implement the MCS we leveraged the SUCHAI mission control software (SUCHAI MCS). This software is a node on the CSP network that has a command line-based interface that allows basic MCS functionality like scheduling telecommands or saving downlinked telemetry. 
Additionally, we provide a desktop environment accessible via VNC that includes Gpredict and the SUCHAI MCS. Finally, we implemented a radio link and network delay simulation using a sleep function whenever the MCS receives data from the space segment simulations.

\textbf{3) \radioname.} The \radioname was implemented as a \emph{ZMQ Hub}~\cite{ProtocolStackCubesatZMQ}. The ZMQ Hub functions as a router for the CSP network simulating both the ground part of the CSP network including the ground station and the space segment. Depending on whether the satellite is currently passing, packets to the space segment are forwarded to the Flight Software Runtime. Additionally, the \radioname dynamically simulates packet loss depending on the predicted elevation angle of the satellite (provided by the \apiname's orbital simulation). For example, if an attacker sends a telecommand, the \radioname will use a probability value based on the elevation angle and decide to drop or forward the packet.


\textbf{4) Ground Configuration.} We implemented the ground configuration as settings that can be modified through a Docker compose file explained in Appendix~\ref{appendix-c}. 

\textbf{5) Logging Repository.} This implementation uses the MongoDB database~\cite{mongodb2024mongodb} to aggregate and store logs. The web interface records login attempts, while the Telnet server logs inputs on a per-connection basis. Additionally, network traffic to and from the host running \honeyname is captured. The logs are secured at points that are not observable by attackers, e.g., different Docker container, shown in Fig.~\ref{fig:docker-arch}, and the isolated MongoDB database is accessed by an append-only user.

\subsubsection{Space Segment Implementation}\label{subsec:space:implementation}

We now describe how we implemented the four space segment simulations designed in Sec.~\ref{subsec:space-design}.
1)~the \emph{Flight Software Runtime}, 
2)~the \emph{Flight Software Services},  
3)~the \emph{Satellite Personality},and 
4)~the~\emph{Logging Repository}.

\textbf{1) Flight Software Runtime.} We use the \emph{SUCHAI Flight Software} (SUCHAI FS)~\cite{gonzalez2019architecture}, which has been deployed in four satellite missions~\cite{Garrido2023} and provides flight software based on the CSP ecosystem.

\textbf{2) Flight Software Services.} The SUCHAI FS uses libCSP~\cite{LibcspLibcsp2024} and thus features the libCSP default services that present a generic attack surface. SUCHAI FS includes a set of telecommands which we modified to use the \apiname instead of querying real sensors.

\textbf{3) Satellite Personality.} The satellite personality was implemented as a Python class that holds multiple variables related to the space segment. It includes almost 50 configurable variables, such as the satellite's name.

\textbf{4) Logging Repository.} 
Like the ground segment's logging repository (Sec~\ref{subsec:ground:implementation}), the \apiname and the FS are connected to the MongoDB instance using a MongoDB client.

Finally, we containerized and hardened this version of the honeypot implementation for secure and easy deployment, which we describe in Appendix~\ref{appendix-c}.

\subsection{Summary of Existing and Newly Developed Software}
\honeyname's implementation involved the integration of existing, modified, and newly developed software. \emph{As-is software} includes TigerVNC, MongoDB, and Gpredict. \emph{Modified software} includes the SUCHAI FS, SUCHAI MCS, the Telnet server, and the ZeroMQ Hub. \emph{Newly developed software} includes the \apiname, including all 6 simulations, the web interface, the flight software services for FS integration, the satellite personality, the ground configuration, and all the necessary framework infrastructure, including Docker files, integration scripts, and configuration files.

%% file: 6-evaluation.tex
\section{Evaluation}\label{sec:evaluation}

In order to evaluate \honeyname we start by listing 
three experimental questions designed to test 
its
alignment with our design objectives. Next, we present four sets of experiments that provide empirical evidence confirming that our design objectives have been addressed as intended. 
We then describe each experiment's environment, methodologies, and results.

\subsection{Experimental Questions}\label{subsec:evaluation:questions}
The following 
questions aim to determine 
if 
\honeyname meets the design objectives outlined in Sec.~\ref{subsec:design:objectives}.
\begin{questions}

    \item \label{question:high-interaction}
    \textbf{Can \honeyname offer extensive interaction opportunities to adversaries?}\\
    Since capturing data on varied techniques is the purpose of any honeypot, we explore the capabilities
    of \honeyname, as described in Sec.~\ref{sec:architecture}.
    This question is related to design objective \ref{objective:rich-interactions} and is addressed in Sec.~\ref{subsec:local-experiments}. 

    \item \label{question:deception}
    \textbf{Can \honeyname simulate a \smallsat mission well enough to deceive adversaries?}\\
    After enticing an adversary, \honeyname must keep its true nature hidden. Thus, \honeyname needs to simulate the satellite's communication and physics characteristics described in Sec.~\ref{subsec:background:satellite-arch}.
    This question relates to design objective \ref{objective:deception} and is answered in Sec.~\ref{subsec:survey} and \ref{subsec:internet-experiment}.

    \item\label{question:extensibility}
    \textbf{Can \honeyname be extended to support different \smallsat{} missions and hardware-in-the-loop simulations?}\\
    \honeyname's customization is important because it would allow users of our framework to implement their own honeypots and even support satellite hardware integration. This question relates to design objective \ref{objective:extensibility}, and we answer it in Sec.~\ref{subsec:case-study} and Sec.~\ref{subsec:hardware-integration}.

\end{questions}

To answer the above research questions, we conducted five experiments. First, in Sec.~\ref{subsec:local-experiments} we craft multiple attacks in a controlled environment to quantify the level of interaction that \honeyname provides. Second, in Sec.~\ref{subsec:survey} we conduct a survey with experienced satellite operators to evaluate \honeyname's realism. Third, in Sec.~\ref{subsec:internet-experiment} we deploy \honeyname and expose it to the Internet to test its deception capabilities. Fourth, in Sec.~\ref{subsec:case-study} we test \honeyname's extensibility by integrating a completely different flight software. Fifth and final, in Sec.~\ref{subsec:hardware-integration} we test \honeyname's support for hardware-in-the-loop operations.

\subsection{TTP Interaction Experiment}\label{subsec:local-experiments}

This experiment seeks to answer \ref{question:high-interaction} by quantifying the interactions provided by \honeyname. To achieve this, we leveraged the SPACE-SHIELD matrix (Version 2.0)~\cite{esa2024space-shield} which provides a collection of adversary tactics and techniques for space systems. We determined the number of tactics and techniques that \honeyname supports. SPACE-SHIELD consists of \tactics and \techniques. However, not all of them are applicable to a virtual, network-based honeypot such as \honeyname.
For example, the technique \emph{Compromise Hardware Supply Chain} involves ``replacing a hardware component in the supply chain with a custom or counterfeit part'' which is out of the scope of a virtual honeypot.
Taking this into consideration,  \finaltactics are applicable to \honeyname. From these tactics, \honeyname can feasibly offer up to 38 techniques as interactions for adversaries.

\textbf{Experiment Description.} Our experimental environment included two hosts. One host running \honeyname and the adversary host. The \honeyname host ran Ubuntu 23.10 and was configured with the SUCHAI-2 satellite and ground personalities. The adversary host ran a Telnet client. Both hosts were connected to the same network.

\textbf{Experiment Methodology.} We designed one interaction or exploit for each of the 38 applicable techniques for our honeypot. The interactions involved ground segment simulations such as the web interface. The exploits were simple, using one TC, or complex with multiple TCs involved. 
For example, the technique \emph{System Service Discovery (ID T1007)} involves adversaries obtaining information about services using tools and OS utility commands. 
Based on this description, we designed the following exploit to capture information about the satellite's running processes:
\lstset{style=TC_commands}

\begin{lstlisting}
1: obc_system ps -aux > ps.log
1: tm_send_file 10 ps.log
\end{lstlisting}

\textbf{Experiment Results.} We successfully crafted and ran an interaction or exploit on \honeyname for 33 techniques out of the feasible 38 as depicted in Table~\ref{table:results:ttps}. The remaining 5 techniques were not implemented due to limitations in  \honeyname's implementation. For example, the \emph{Retrieve TT\&C master/session keys (T2015.002)} technique depends on a cryptographic protocol implementation which \honeyname's flight software currently does not support. Due to space limitations, we do not describe all the exploits and interactions here. However, the complete exploit and interaction list is available in Table~\ref{table:exploits} in Appendix~\ref{appendix:exploits}. 

The key findings of our experiment are shown below, providing 
strong evidence for answering question~\ref{question:high-interaction} in the affirmative and design objective~\ref{objective:rich-interactions} as achieved.

\begin{myBox}[]{Key findings \ref{question:high-interaction}}{}

\begin{itemize}[nosep,leftmargin=0pt,labelindent=0pt]
    \item \honeyname supports 87\% of the SPACE-SHIELD matrix techniques possible in a virtual satellite honeypot.
    \item \honeyname supports 100\% of the SPACE-SHIELD matrix tactics.
\end{itemize}

\end{myBox}

\input{tables/ttps-results}

\input{11-survey-evaluation}

\subsection{Internet Interaction Experiment}\label{subsec:internet-experiment}

This experiment explores \honeyname's capabilities to entice external actors by deploying it over the Internet.

\textbf{Experiment Methodology.} 
We leveraged \honeyname's customization and deployability features to deploy five instances of \honeyname over the Internet and exposed TCP ports 23 for Telnet, 80 for the web interface, and 5901 for VNC.

\textbf{Deployment.} Initially, we deployed our first \honeyname instance to simulate the SUCHAI-2 \smallsat on the premises of the University of Chile, where the mission's real ground segment is. This on-premises deployment provides the most convincing environment to adversaries.  After coordinating with the University of Chile's local stakeholders for three months, we deployed \honeyname on an onsite server using a University of Chile IP address starting on July 2024.

Although we went to great lengths to make our deployment as realistic as possible, after a few months, our honeypot did not entice any external actors.
Therefore, in January 2025, we deployed four additional \honeyname instances on cloud servers. We configured two instances to simulate NASA's Advanced Composite Solar Sail System (ACS3) CubeSat~\cite{nasa2025acs3}, and two simulating DLR's PIXL-1 CubeSat~\cite{eoportal2025pixl1}. We selected these two particular CubeSats because they use the CSP ecosystem (Sec.~\ref{subsec:ecosystems}) and are currently in-orbit and operational.
We ensured these additional deployments were believable by configuring their satellite personalities and giving them public IP addresses in the regions belonging to the satellite mission origin, namely Germany and the United States. Table~\ref{tab:deployment_summary} summarizes all five \honeyname deployment details.

While IP assignment can affect our honeypots' covert status~\cite{lopez2022honeyplc}, the increasing use of cloud and web-based ground segment services, such as AWS Ground Station~\cite{amazon2024satellite} and YAMCS~\cite{yamcs}, has made cloud-hosted satellite operations increasingly commonplace. As a result, a cloud deployment no longer inherently signals inauthenticity to adversaries. Furthermore, the part of the Ground Segment that we expose on cloud IPs is not dependent on being collocated with the GS and thus may as well be hosted with a cloud provider.

\input{tables/deployment-details}

\input{tables/table-internet-interactions}



\textbf{Experiment Results.} We collected several gigabytes of network traffic data on ports 23, 80 and 5901 for each honeypot, however most of these data was generated by crawlers and bots and not satellite-specific. But, on January 18th, 2025, we received our first space-specific interaction via Telnet and in the next few months we received more. Overall, three of our honeypots successfully enticed external actors and captured four distinct interaction sessions (shown in Table~\ref{table:telnet-interactions}) all via Telnet.
The four interaction sessions yielded 22 flight software-specific commands. These flight software-specific commands show that the adversaries that interacted with \honeyname were purposefully trying to exploit our honeypot using commands that the flight software recognized.

Across the four Internet interaction sessions we observed, the first lasted two hours
and originated from a non-Tor IP in Egypt. The next two sessions occurred within a week,
shared their command set (e.g.,  \texttt{help}, \texttt{ls}, and \texttt{fp\_show}), and transitioned to Tor exit nodes in Sweden and France. These three factors suggest that the same adversary returned using anonymized infrastructure after initial reconnaissance. The first session may have ended after the actor gathered sufficient information. 
In contrast, the fourth and final interaction in April, separated by over two months, from a non-Tor U.S. address, and using a very different command set appears unrelated and likely
originates from a different actor. Overall, the first three interactions are best interpreted as a single
adversary adapting its anonymity strategy, whereas the last session was likely initiated
by a second actor.

We now describe one of the most complex interaction sessions \honeyname captured and analyze it in terms of the SPACE-SHIELD matrix (discussed in Sec.~\ref{subsec:ttps}). In this session, the adversary targeted one of the ACS3 honeypots located in the United States. First, they connected to the exposed Telnet server thus getting access to the ground segment, this corresponds to the \emph{Ground Segment Compromise (ID T1584.001)} technique. Then they issued the \emph{``help''} and \emph{``test''} commands which list the available commands supported by the MCS and flight software; these commands correspond to the \emph{Gather Victim Mission Information (ID T2002)} technique. 
Later, the adversary issued the \emph{``tm dump 0x0000''}, which writes telemetry data to a file, showing an attempt to extract mission data, which matches the \emph{Exfiltration Over TM Channel (ID T2022)} technique. Next, they sent the \emph{``com\_ping''} command which attempts to initiate an interaction with the target spacecraft, matching the \emph{Active Scanning (RF/Optical) (ID T2001)} technique.
The adversary again implemented the \emph{Exfiltration Over TM Channel (ID T2022)} technique by issuing the \emph{``tm\_parse\_beacon''} command which prints beacon telemetry (lightweight periodic telemetry) and the \emph{``obc\_get\_sensor''} command which instructs the OBC to print a particular sensor's data.
Finally, the adversary attempted to tamper with the spacecraft's OBC by issuing the \emph{``obc\_update\_status''} command. This command updates the OBC status variables to any arbitrary value, thus matching the \emph{Modification of On Board Control Procedures (ID T2010)} technique.

Overall, this interaction session alone involved a chain of five SPACE-SHIELD matrix techniques. However, after analyzing all four interaction sessions, we identified the \emph{Spacecraft's Components Discovery (ID T2034)} technique, totaling six SPACE-SHIELD techniques based on real-world data captured by \honeyname after successfully enticing and deceiving human adversaries.

While our dataset is relatively small, its significance becomes evident when contextualized. For example, according to a 2018 Open Platform Communications (OPC) group report~\cite{arc2018opcinstalledbase}, there are an estimated 47 million OPC-enabled ICS devices deployed worldwide, compared to 11,000 operational satellites in 2025~\cite{livescience2024satellites}. This represents about \emph{4,700 ICS devices for every one satellite}. In other words, attackers have orders of magnitude more opportunities to target ICS devices than satellites. Thus, making our dataset a rare and valuable contribution to understanding space systems' TTPs.

In addition to its rarity, our dataset is the \emph{first} empirical real-world dataset of TTPs against satellites. As such, it not only provides the first view into how adversaries interact with satellite systems, but also establishes a reproducible baseline reference for future studies. Lastly, our contribution complements prior work that provides limited insights, e.g., no specific telecommands~\cite{uscc2011annualreport}.

In summary, these results provide strong evidence
for answering question \ref{question:deception} in the affirmative and design
objective \ref{objective:deception} as achieved.

\begin{myBox}[]{Key findings \ref{question:deception}}{}


\begin{itemize}[nosep,leftmargin=0pt,labelindent=0pt]
    \item \honeyname deceived human adversaries who sent \emph{22 satellite flight software-specific commands}.
    \item The interactions collected by our \honeyname deployments comprise six SPACE-SHIELD techniques.
\end{itemize}

\end{myBox}

\subsection{Case Study: Extending \honeyname  to  CCSDS Ecosystem}\label{subsec:case-study}

In Sec.~\ref{subsec:cspgeneric} we described how we implemented \honeyname using one \smallsat protocol ecosystem, namely, CSP. In this case study, we are interested in testing the extensibility capabilities of
\honeyname to support additional ecosystems (discussed in Sec.~\ref{subsec:ecosystems}) by adding a \emph{second ecosystem} to \honeyname, namely the CCSDS ecosystem.

\textbf{Experiment Description.} We selected CCSDS because it is a standard protocol suite used by other \smallsat{}s~\cite{willbold2023space}. To accomplish this, we leveraged an open-source CCSDS ecosystem-based flight software framework, RACCOON OS~\cite{raccoonos}, and YAMCS, an open-source Mission Control software framework with built-in support for PUS~\cite{yamcs}.

\textbf{Experiment Methodology.} Building upon our \honeyname framework implementation, as detailed in Sec.~\ref{sec:implementation}, we enhanced the system by integrating various components of the RACCOON OS and the YAMCS framework.

Regarding the ground segment, we implemented the \emph{exposed network protocol} using YAMCS' built-in web interface. For the \emph{mission control software}, we used YAMCS's built-in Mission Control Software~\cite{yamcs}. For the \emph{radio simulator}, we used RACCOON's communication application. For both the \emph{ground configuration} and the \emph{logging repository}, we again used YAMCS built-in features.

For the space segment, we implemented the \emph{flight software runtime} using the RACCOON framework. For the \emph{flight software services}, we configured the RACCOON framework to connect it to the YAMCS's MCS on the ground segment. The \emph{satellite personality} and \emph{logging repository} were based on the existing \honeyname implementations.

\textbf{Experiment Results.} We successfully extended \honeyname to support the CCSDS ecosystem. The implementation was completed by a graduate student with no prior satellite-related experience in the span of two weeks. However, the majority of the effort involved in extending \honeyname to support the CCSDS ecosystem was dedicated to understanding the ecosystem itself, and analyzing the RACCOON flight software code and the YAMCS framework documentation. The only extra implementation that was required was the flight software services which we modified using Rust. Other than that, we reused several  modules such as the \apiname. 

In summary, these results provide evidence
for answering \ref{question:extensibility} in the affirmative and \ref{objective:extensibility} as achieved.

\begin{myBox}[]{Key findings \ref{question:extensibility}}{}

Out of the box \honeyname supports CSP and CCSDS, the two most widely used space ecosystems, and was evaluated by simulating \emph{3 real-world \smallsat{}s}.


\end{myBox}

\input{tables/supported-sats}

\subsection{Case Study: Hardware-in-the-loop Experiment}\label{subsec:hardware-integration}

This experiment is designed to 
produce evidence of the robustness of 
\honeyname
to support real-world satellite hardware-in-the-loop (HIL) operations. 

\textbf{Experiment Description.}
In this experiment we integrated \honeyname with an \emph{in-orbit, operational \smallsat} mission. To achieve this, we collaborated with an aerospace company that develops hardware subsystems for \smallsat{s}. The experiment involved sending a telecommand from \honeyname's ground segment simulation which was then sent to the in-orbit \smallsat, which would then process 
the telecommand. Due to security restrictions, we are unable to disclose the name of the in-orbit \smallsat which we refer to as \smallsat X.

\textbf{Experiment Methodology.} In order to achieve the HIL integration, we 
customized 
both \honeyname and \smallsat X's mission. 
For \honeyname, 
we modified the \apiname to support \smallsat X's mission control software (YAMCS). Specifically, we deployed a proxy server that connected the \apiname to the \smallsat X's mission production YAMCS instance.
Conversely, the 
\smallsat X's environment was customized by aerospace company's team by deploying a script in their YAMCS instance to complete the integration.
Once the integration was completed, we coordinated with the aerospace company to send a telecommand from \honeyname during one of \smallsat X's passes.



\textbf{Experiment Results.} The telecommand successfully reached \smallsat X during one of the passes and \honeyname received the appropriate telemetry.
Due to security restrictions, we are not able to disclose details on the telecommand and the telemetry received. However, the entire communication was initiated by \honeyname which in turn received sanitized telemetry from \smallsat X, effectively closing the loop in the HIL experiment. Additionally, the aerospace company provided us with a snippet of radio signal data shown in Fig.~\ref{fig:waterfall}, which confirms that the experiment was successful.
In summary, these results highlight \honeyname's robust extensibility features and provide strong evidence
for answering question \ref{question:extensibility} in the affirmative and design
objective \ref{objective:extensibility} as achieved. These findings serve as empirical evidence of \honeyname's fidelity when interacting with a real satellite mission.

\begin{myBox}[]{Key findings \ref{question:extensibility}}{}

    \honeyname was successfully integrated into a satellite hardware-in-the-loop operation and communicated with an \emph{in-orbit, operational \smallsat}.

\end{myBox}

%% file: tables/ttps-results.tex
\begin{table}[t]
    \centering
    \caption{Tactics and techniques supported by \honeyname.}
    \scriptsize

    \begin{tabular}{lcc}
    \toprule
      Tactics & \makecell{SPACE-SHIELD\\ Techniques\\(Applicable to Virtual Honeypots)} & \makecell{\honeyname \\ Supported \\Techniques}\\
      
      \toprule
        Reconnaissance  & 2 & 2  \\ \hline
        \makecell[l]{Resource Development} & 2 & 2 \\ \hline
        \makecell[l]{Initial Access} & 2 & 2 \\ \hline
        Execution & 2 & 2 \\ \hline
        Persistence & 2 & 2 \\ \hline
        \makecell[l]{Privilege Escalation} & 2 & 1 \\ \hline
        \makecell[l]{Defense Evasion} & 4 & 4 \\ \hline
        \makecell[l]{Credential Access} & 3 & 3 \\ \hline
        Discovery & 2 & 2 \\ \hline
        \makecell[l]{Lateral Movement} & 4 & 1 \\ \hline
        Collection & 2 & 2 \\ \hline
        \makecell[l]{Command \& Control} & 2 & 2 \\ \hline
        Exfiltration & 2 & 1 \\ \hline
        Impact & 7 & 7 \\ \hline
        \toprule
        Total & 38 & 33 \\
        
        \bottomrule
    \end{tabular}
    \label{table:results:ttps}
\end{table}

%% file: 11-survey-evaluation.tex
\subsection{\smallsat Operators Survey}\label{subsec:survey}

Evaluating the realism of a satellite honeypot is challenging for two main reasons. First, as discussed in Sec.~\ref{sec:introduction}, satellites, including \smallsat{}s, are very diverse. Second, there is no established metric or tool, such as Nmap's OS detection~\cite{nmap2025nmap}, to quantify the level of realism of our honeypot.

To evaluate the realism and deception capabilities of \honeyname, we surveyed \emph{experienced \smallsat operators}.
Because operators interact with real-world \smallsat missions on a daily basis they are \emph{experts} and thus are the best population to rigorously evaluate \honeyname. 

\input{survey-figures/question-types-table}

\subsubsection{Survey Structure}
We divided our survey into \surveysections sections. The first section collected the participants' background information, e.g., demographic data, the second section focused on their professional experience, the third section probed participants’ satellite operation experience, e.g., how many missions they had operated.

The fourth survey section involved giving participants \emph{hands-on access} to a \emph{live instance of \honeyname}. In this section, participants were asked to perform 4 hands-on tasks that correspond to specific satellite mission operations using a live instance of \honeyname as depicted in Fig.~\ref{fig:suchai-vnc-honeypot}.
Each hands-on task was designed to feature different components of \honeyname by replicating a real-world satellite mission operation, as discussed in Section~\ref{sec:background}. After participants completed each hands-on task, they answered questions designed to assess how realistic they found different aspects of the honeypot. Table~\ref{table:surveyquestions} lists the satellite operations included in the survey.

Finally, in section five, once participants had interacted with multiple elements of our honeypot, the survey concluded with participants answering questions about \honeyname's overall realism and deception. Questions related to both the honeypot operation tasks and the overall evaluation used a 5-point Likert scale~\cite{likert2025qualtrics} to measure the participants' reactions.

\subsubsection{Participants}
We conducted the survey via Qualtrics and Zoom videoconferencing and distributed the survey directly to operators from previously identified missions.
In total, we received responses from \operators satellite operators who have between 1 to 10 years of experience operating satellites and have operated between 1 to 5 unique missions.
In terms of demographics, 20\% (2/10) of the participants were female and 80\% (8/10) male. 30\% (3/10)  belonged to the 18-24 age group, 40\% (4/10) to the 25-34 age group, and 30\% (3/10) to the 35-44 age group. 
In regards to geographic location, 40\% (4/10) of participants were located in Europe, 20\% (2/10) in North America and 40\% (4/10) in South America.





Recruiting participants was challenging due to the rarity and specialized nature of the required expertise. Over a span of four months, we reached out to national and international institutions, as well as private corporations involved in operational satellite missions to identify suitable participants. Despite these challenges, our sample is diverse, including operators from Europe, North America, and South America, spanning industry, government, and academia, across 5 missions.

\subsubsection{Methodology and Key Results}

In the survey, we evaluated three key aspects of our honeypot. First, whether the ground segment simulations are realistic; second, whether the space segment simulations are realistic; and third, whether \honeyname, as a whole, provides a convincing and realistic \smallsat mission simulation.

Before describing the results, it is important to emphasize that the participants were \emph{informed} that the system they interacted with was a simulation. We informed participants for two reasons. First, it was not feasible to synchronize participants' availability to take the survey with the timing of the real satellite's pass, which may happen only a few times per day and lasts only a few minutes. Second, claiming to provide access to a real satellite would itself be seen as unrealistic.

\textbf{Ground Segment Realism.} To understand if \honeyname's ground segment is realistic, participants interacted with \honeyname by performing different satellite operations (discussed in Sec.~\ref{subsec:satellite-context}), to showcase the ground segment components listed in Table~\ref{table:surveyquestions}. 
One of these tasks is the \emph{pass prediction} operation which involves calculating when and where a satellite will be within the communication range of a specific ground station. This operation is performed using the ground station control software. After the participants performed this operation on \honeyname, we asked them to rate their perceived level of realism. 90\% of the participants strongly agreed and 10\% agreed that the \emph{pass prediction} operation they performed on \honeyname resembles that of a real mission.
%

Another relevant operation is the \emph{telecommand scheduling} operation which involves the planning and queuing of the commands to be sent to the satellite during a pass. Again, after performing this operation on \honeyname, 90\% of the participants strongly agreed and 10\% agreed that the telecommand scheduling operation performed resembles that of a real mission mentioning the use of a command line interface-based mission control software as a contributing factor.

In summary, according to these results, the vast majority of participants, who are \emph{experienced satellite operators}, perceive the ground segment simulated by \honeyname as highly resembling a real satellite mission.

\textbf{Space Segment Realism.} Similarly, to understand \honeyname's space segment realism level, we gave participants hands-on access to a live instance of \honeyname and asked them to perform different operations that make use of \honeyname's space segment simulations (discussed in Sec.~\ref{subsec:space-design}).

For example in the \emph{telemetry download} operation, participants issued multiple telecommands to \honeyname's space segment simulation to download a plethora of telemetry data, including EPS, temperature and ADCS data, which was generated in real time by our \apiname.


After the participants performed the telemetry download operation, we asked them to rate their perceived level of realism of the telemetry output shown during the operation. 90\% of the participants agreed or strongly agreed that the telemetry shown during the live \emph{telemetry download} operation resembles that of a real mission, with the temperature telemetry achieving a 100\% strongly agree response rate with some participants mentioning that the temperature values resemble a normal operation and align with the expected temperature values of a satellite. 
These results indicate that the telemetry generated by the \apiname is considered highly realistic by the vast majority of participants.

In summary, according to these results, the vast majority of participants, who are \emph{experienced satellite operators}, perceive the space segment simulated by \honeyname as highly resembling a real satellite mission.

\textbf{\honeyname's Overall Realism.} Finally, to understand \honeyname's overall realism, we asked the participants a series of overall evaluation questions after they had interacted with \honeyname. For example, we asked participants to evaluate the realism of the telecommands used in all hands-on tasks. 70\% strongly agreed, 20\% agreed and 10\% neither agreed nor disagreed that \honeyname's telecommands are realistic. Additionally, when asked if they would be able to distinguish between \honeyname's satellite mission simulation and a real mission, 70\% strongly agreed, 20\% agreed and 10\% neither agreed nor disagreed that they would \emph{not} be able to and again to emphasize that these are \emph{experienced satellite operators}.

Overall, our survey results provide evidence for answering question~\ref{question:deception} in the affirmative and design objective~\ref{objective:deception} as achieved. These findings serve as empirical evidence of \honeyname's fidelity when simulating a real satellite mission.
Our survey's questions and results are available online\footnote{\url{https://github.com/HoneySat/honeysat-survey-data}.}.

\begin{myBox}[]{Key finding \ref{question:deception}}{}
90\% of surveyed satellite operators agreed with the statement: \emph{``I~would not be able to distinguish between \honeyname's satellite honeypot system from the real CubeSat satellite mission it is based on.''}

\end{myBox}

%% file: survey-figures/question-types-table.tex
\setlength{\tabcolsep}{5pt}
\begin{table}[b]
\footnotesize
\centering
\caption{Summary of questions in Section~\ref{subsec:survey}, satellite honeypot operation tasks of the survey.}
\begin{tabular}{c|c|c}
\toprule
\textbf{Satellite Operation} & \textbf{Evaluated Component} &
\textbf{No. Qts.}  \\
\midrule
Telecommand Scheduling  & Mission Control SW & 5 \\
\hline
Pass Prediction & Ground Station Control SW & 5 \\
\hline
Telemetry Download & \apiname & 7 \\
\hline
Ping Test & Radio Simulator & 5 \\
\bottomrule
\end{tabular}
\label{table:surveyquestions}
\end{table}

%% file: tables/deployment-details.tex
\setlength{\tabcolsep}{6pt}
\begin{table}[t]
\centering
\caption{\honeyname Internet deployments' details.}
\begin{tabular}{@{}c c c c@{}}
\toprule
\textbf{\begin{tabular}{@{}c@{}}Deployment \\ Type\end{tabular}} & 
\textbf{\begin{tabular}{@{}c@{}}Satellite \\ Personality\end{tabular}} &  
\textbf{\begin{tabular}{@{}c@{}}\honeyname IP \\ Location\end{tabular}} & 
\textbf{\begin{tabular}{@{}c@{}}Duration \\ (months)\end{tabular}} \\
\midrule
Cloud     & PIXL-1 & Germany & 6 \\
Cloud     & PIXL-1 & Germany & 6 \\
Cloud     & ACS3 & USA & 6 \\
Cloud     & ACS3 & USA & 6 \\
On-prem   & SUCHAI-2 & University of Chile & 12 \\
\bottomrule
\end{tabular}
\label{tab:deployment_summary}
\end{table}

%% file: tables/table-internet-interactions.tex
\setlength{\tabcolsep}{2pt}
\begin{table}[t]
    \centering
    \caption{Exposed Telnet interactions received.} 
    \begin{tabular}{lccccc}
    \toprule
      \textbf{\begin{tabular}[c]{@{}c@{}}Date\end{tabular}} &
      \textbf{\begin{tabular}[c]{@{}c@{}}Satellite \\Personality\end{tabular}} &
      \textbf{\begin{tabular}[c]{@{}c@{}}\honeyname \\ IP\end{tabular}} &
      \textbf{\begin{tabular}[c]{@{}c@{}}Attacker \\ IP\end{tabular}} &
      \textbf{\begin{tabular}[c]{@{}c@{}}Cmds\\Received\end{tabular}} &
      \textbf{\begin{tabular}[c]{@{}c@{}}Time\end{tabular}} \\
    \toprule
      Jan 18, 2025  & ACS3  & USA  & Egypt         & 4 & 2 hr  \\
      Jan 24, 2025  & PIXL-1 & Germany        & Sweden (Tor)        & 6 & 4 min \\
      Jan 24, 2025  & ACS3   & USA  & France (Tor)        & 9 & 5 min \\
      Apr 3, 2025   & ACS3   & USA  & USA  & 8 & 3 min \\
    \bottomrule
    \end{tabular}
    \label{table:telnet-interactions}
\end{table}

%% file: tables/supported-sats.tex
\setlength{\tabcolsep}{6pt}
\begin{table}[t]
\centering
\caption{\smallsat ecosystems (2) and example \smallsat{}s (3) integrated to \honeyname.}
\begin{tabular}{@{}c c c c@{}}
\toprule
\textbf{\begin{tabular}{@{}c@{}}Honeypot \\ Framework\end{tabular}}& 

\textbf{\begin{tabular}{@{}c@{}}Generic \\ \smallsat\end{tabular}}&  

\textbf{\begin{tabular}{@{}c@{}}Ground \\ Configuration\end{tabular}} &
\makecell{\textbf{Institution}} \\
\midrule
HoneySat & CSP   & \makecell{ACS3}     & NASA \\ 
         &       & \makecell{PIXL-1}      & DLR \\ 
         &       & \makecell{SUCHAI-2} & U of Chile \\ 
\cmidrule(lr){2-4} 
         & CCSDS & \makecell{ PUS}   & N/A \\ 
\bottomrule
\end{tabular}
\label{tab:honeysat}
\end{table}

%% file: 7-discussion.tex
\section{Discussion and Future Work}

\textbf{Challenges of Creating the First Satellite Honeypot.} During \honeyname's design and implementation we encountered and overcame two main challenges which stemmed from fundamental differences between satellites and other Cyber-Physical Systems (CPS), e.g., ICS.

First, \emph{time and link-constrained communication.}  Unlike ICS honeypots, which assume continuous network connectivity and stable control loops, e.g., scan cycles, satellite communication occurs only during orbital passes with variable duration and packet loss. To overcome this challenge, we developed the \radioname discussed in Sec.~\ref{subsec:ground-design}.

Second, \textit{ dynamic physics-aware simulation.}  
ICS honeypots simulate sensor readings through simple physics loops (e.g., tank level, valve position). In contrast, a satellite honeypot must dynamically maintain  physically coherent telemetry across interdependent subsystems such as attitude, and power. We addressed this by creating the \textit{Satellite Simulator}, which simulates subsystems and sensors that communicate between each other and generate data in real time.

\textbf{Satellite Honeynets.} To increase attacker engagement and reflect emerging satellite networks~\cite{ma2023network}, multiple \honeyname instances can be deployed as a honeynet~\cite{icsnet2024salazar}. Each instance could simulate different satellites and communicate through virtual inter-satellite links (ISLs). In addition, \honeyname could also be part of a honeynet connected to other ground mission infrastructure such as database servers and radio equipment. Due to \honeyname's extensibility, these honeynet scenarios could be implemented in the future.

\textbf{Satellite Honeypots' Fingerprinting.}
Based on our experience designing, implementing and evaluating \honeyname we now discuss some qualitative detection vectors that can guide future satellite honeypot fingerprinting research.

\emph{Space Protocol Uniformity.} 
The use of default space protocol configurations may be used for fingerprinting. For example, CSP deployments built directly from libCSP expose identical port assignments and diagnostic services, e.g. ping.

\emph{Perfectly periodic pass timing.}  
Real satellites rarely follow identical pass schedules; orbital perturbations, TLE drift, and operational delays introduce small timing variations. A honeypot that enables communication at perfectly fixed intervals or with constant latency can thus be fingerprinted. To avoid this, \honeyname periodically downloads the latest TLE data to keep the orbital simulation from drifting. This, on the other hand, causes a jump in the orbit once the new TLE data is used to base the simulation on.

\emph{Telecommand implementation.} A given FS may have completely different telecommand names and functions depending on the satellite and thus a mismatch may reveal the honeypot. Additionally, telecommands may have revealing features that print information about the underlying honeypot host, e.g., Ubuntu, which may reveal the simulation.

\subsection{Initial Satellite Honeypot Anti-Detection Techniques}

Building upon lessons learned from \honeyname's development and from other CPS honeypots~\cite{uitto2017survey,tay2023taxonomy}, we identify concrete strategies to enhance \honeyname's and future satellite honeypots' resistance to fingerprinting and detection.

\paragraph{Improve Environmental Diversity}
Previous research shows that static structural layouts and repeated host fingerprints enable easy detection~\cite{tay2023taxonomy}. Future satellite honeypots and honeynets can incorporate diversity in its ground segment topology, and network services. Following the Purdue-style segmentation~\cite{298052} used in ICS honeypots, different \honeyname modules (mission control, station control, payload) can be isolated behind realistic network layers such as routers and firewalls.

\paragraph{Implement Adaptive Reconfiguration}
Honeypot adaptive reconfiguration occurs when a honeypot recognizes that it is being probed or scanned, e.g., Nmap probes, and reconfigures itself to avoid detection~\cite{gabrys2024honeyganpotsdeeplearning,uitto2017survey}.
Adapting this technique to satellite honeypots would involve both the ground and space segments. For example, the honeypot could modify exposed services, e.g., CSP over ZeroMQ to appear as a different ground station node. Additionally, different telemetry datasets could be rotated, or switched between low interaction and high interaction versions of the honeypot~\cite{kyung2017honeyproxy}.

\paragraph{Use Real hardware and software identifiers}
Future satellite honeypots could use real hardware identity and fingerprints such as MAC Organizationally Unique Identifiers (OUIs), and telemetry headers such as APIDs (Application Process Identifiers)~\cite{SpacePacketProtocol2020}. Obtaining these details is non-trivial as they are often proprietary or absent from public documentation, and may require reverse-engineering, or careful analysis of captured mission traffic.

\textbf{Limitations.} Currently, the functionality of some subsystems within \honeyname's \apiname are constrained by the quality of the data provided. For example, the resolution of Earth's images generated by our camera payload depends on the resolution of its source data (USGS~\cite{geological2022earthexplorer}). Consequently, creating a honeypot for a satellite with a high-resolution camera payload would not be feasible without addressing the underlying issues of data source quality.

\textbf{Broader Applications of \honeyname}. Originally designed as a honeypot, our framework can potentially support a range of applications beyond its initial purpose. One promising use case is the development of digital twins for satellite systems, enabling the simulation of real-world satellite subsystems and communication scenarios. Furthermore, \honeyname can be integrated into cyber range environments to enhance cybersecurity training exercises.

%% file: 8-conclusion.tex
\section{Conclusion}
Although we have yet to witness a Stuxnet-like cyberattack on space systems, security researchers need to develop effective countermeasures to secure satellites. In this paper, we introduced \honeyname, the first satellite honeypot that simulates a satellite mission and provided evidence that our honeypot can obtain real-world interaction data and can be extended with real-world satellite software and hardware.
Finally, we hope that security researchers use \honeyname's open-source implementation as a foundation not only for satellite honeypot deployments but also for simulation, and training applications.

\section*{Acknowledgments}
We thank our shepherd and the anonymous reviewers for their helpful suggestions towards improving this paper. The project underlying this paper was funded by the Federal Ministry of Transport (BMV) under the code 45AVF5A011 and was partially supported by NSF awards No. 2131263 and 2232911, and by the US Department of Transportation (USDOT) CYBER-CARE Grant No. 69A3552348332. We further thank the Saarbrücken Graduate School of Computer Science for their funding and support. The authors are responsible for the content of this publication.

\newpage

%% file: ethics-and-open-science.tex
\clearpage

\section{Ethical Considerations}

In this paper, we consider the ethical consequences and possible negative outcomes of the satellite operator user study and the \honeyname deployments.
We now discuss the stakeholders, potential risks, and how we mitigated those risks.

\subsection{Satellite Operators User Study}

Prior to conducting any research, the survey protocol, and materials were submitted for review by our Institutional Review Board (IRB). The protocol received ``exempt status'', indicating no more than minimal risk.

\textbf{Stakeholder Identification.} The primary stakeholders in the survey are \emph{survey participants}.

\textbf{Risk Mitigation.} The main risk for the survey participants is the breach of privacy. To mitigate this risk, we ensured that no identifiable data (e.g., names) was collected, and the responses remain anonymous. Additionally, the survey itself includes an informed consent section that informs the respondents about the voluntary participation, the survey’s purpose, a description of the procedures, the risks involved, contact information, and the option to opt out of the survey. Finally, we provided participants the choice to skip questions using the response options  ``Prefer not to say'' and ``I do not know.''

\subsection{Honeypot Deployment Experiment}

\textbf{Stakeholder Identification.} The primary stakeholders in the honeypot deployment are the \emph{external actors} who interact with our honeypot and the \emph{infrastructure owners} on whose infrastructure our honeypot is running.

\textbf{Risk Mitigation.} The main risk for the infrastructure owners is that external actors will use our honeypot as a stepping stone to breach their infrastructure. For the on-prem deployments, we worked with the local IT administrators to ensure that we took all the necessary precautions to avoid this scenario. First, we were provided with a server with a clean OS installation that did not have any production applications or sensitive data. Second, the server was isolated on its own network segment. Third, we configured a firewall to only allow the necessary ports. Fourth, the VNC portion of the honeypot is configured so that external actors cannot use it to host malicious services by denying traffic forwarding. Fifth and final, we deployed our honeypot using two sandboxing layers (Docker containers and Virtual Machines (VMs) as we discussed in Appendix~\ref{appendix-c}. 

The main risk for the external actors is the breach of their privacy. However, it is generally accepted that trespassers of a computer system do not have reasonable expectation of privacy~\cite{spitzner2002honeypots}. In order to mitigate this risk, our honeypot system gives notice and warning to any user connecting to it indicating that ``This computer system is for authorized use only.'' Additionally, the web services of our honeypot are protected by a username and password.

%% file: 9-appendix.tex

\section{Deployment and Security Hardening Implementation}\label{appendix-c}

\begin{figure}[ht]
    \centering
    \includegraphics[width=1.0\linewidth]{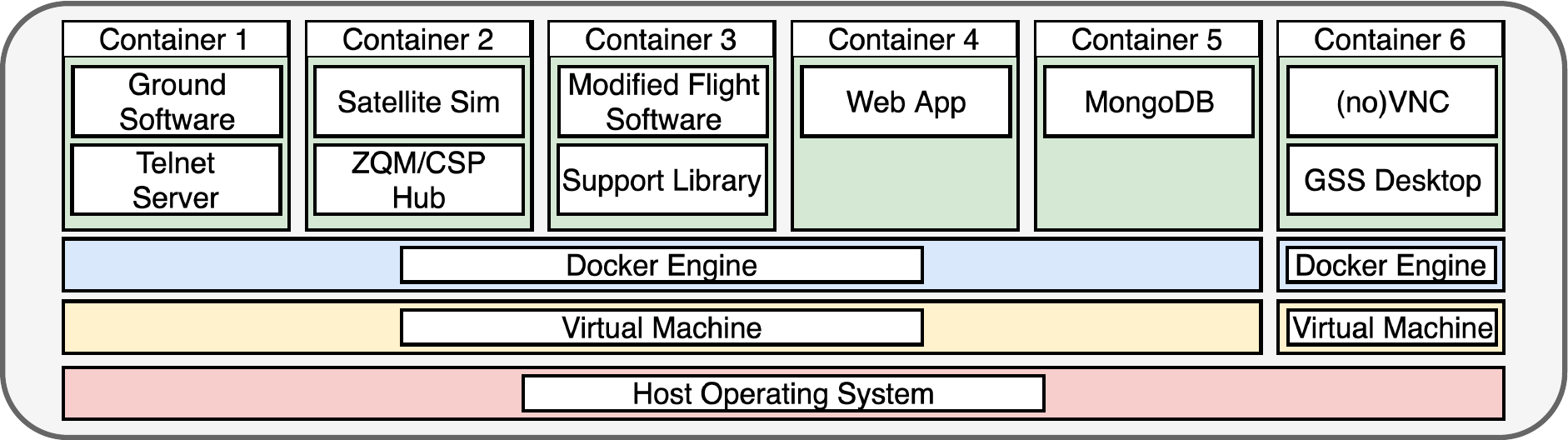}
    \caption{Architecture of dockerized Generic CSP Honeypot}
    \label{fig:docker-arch}
\end{figure}

As we mentioned in Sec.~\ref{subsec:honeypots}, high-interaction honeypots such as \honeyname present a high risk of adversary takeover. To mitigate this risk, we implemented \honeyname with two sandboxing layers.

\textbf{Virtual Machine.} VMs provide the highest isolation
level among sandboxing techniques. We implemented \honeyname in VM environments to leverage VMs' robust security.

\textbf{Containerization.} After we completed \honeyname's development, we used Docker Compose~\cite{docker2024docker} to containerize each of our framework's applications. Specifically, we created four different containers depicted in Fig.~\ref{fig:docker-arch}.
Containerizing \honeyname provides two benefits. First, it creates another sandboxing layer that prevents adversaries from using our honeypot to access the underlying system~\cite{marra2024feasibility}. Second, it proves a convenient and flexible way to deploy \honeyname.

\input{tables/comparison-real-sim-satellite}

\section{Interaction Sequences and Exploits}\label{appendix:exploits}

Table~\ref{table:exploits} includes all the interaction sequences and exploits we performed during the experiments in Sec.~\ref{subsec:local-experiments}.
\input{tables/table-exploits}

\section{Hardware-in-the-loop Experiment Results}
\begin{figure}[h]
    \centering
    \includegraphics[width=0.95\linewidth]{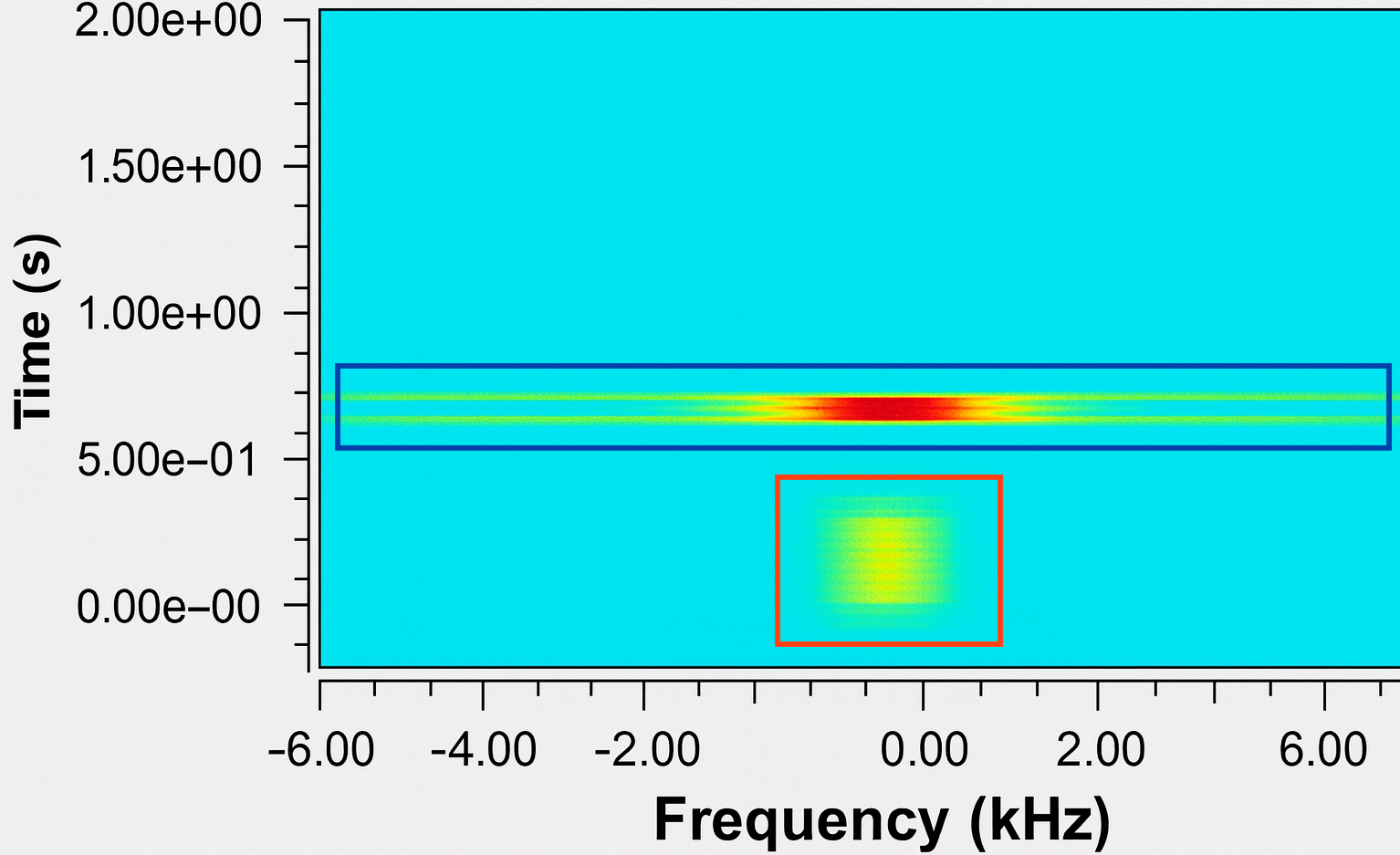}
    \caption{Waterfall plot that shows the radio signal spectrogram when \honeyname communicated with \smallsat X. The signal on top (blue box) belongs to the telecommand sent by \honeyname, and the signal on the bottom (red box) belongs to the telemetry received from the \smallsat X.}
    \label{fig:waterfall}
\end{figure}

Fig.~\ref{fig:waterfall} depicts a snippet of radio signal data which confirms that the hardware-in-the-loop experiment in Sec~\ref{subsec:hardware-integration} was successful. This was provided by the aerospace company.

%% file: tables/comparison-real-sim-satellite.tex
\setlength{\tabcolsep}{4.5pt}
\begin{table}
\footnotesize
\centering
\caption{Correspondence between real satellite mission components and \honeyname's design components.}
\begin{tabular}{c|c|c}
\toprule
\textbf{Segment} & \textbf{Real Component} & \textbf{HoneySat Component} \\
\midrule
Ground & Remote Operation Protocols & VNC, Telnet \\
\hline
Ground & Mission Control Software & Modified MCS \\
\hline
Ground & Ground Station Control Software & Gpredict \\
\hline
Ground & Web Interface & Web Interface \\
\hline
Ground & Ground Station & Radio Simulator \\
\hline
Space & Flight Software & F.S. Runtime + Services \\
\hline
Space & On Board Computer & Docker Container \\
\hline
Space & Platform & \apiname{} \\
\hline
Space & Payload & \apiname{} \\
\bottomrule
\end{tabular}
\label{table:real-components-comparison}
\end{table}

%% file: tables/table-exploits.tex

\begin{table*}[]
\centering
\caption{Tactics, Techniques and Procedures' Interaction Experimental Exploits.}
\label{table:exploits}
\resizebox{\textwidth}{0.50\textheight}{%
\begin{tabular}{lllll}
\toprule
\textbf{Tactic} &
  \textbf{Technique} &
  \textbf{ID} &
  \textbf{Subsystem} &
  \textbf{Exploit} \\ \midrule
Reconnaissance &
  Active Scanning (RF/Optical) &
  T2001 &
  Threat Model Limitation &
  N.A. \\
Reconnaissance &
  Gather Victim Mission Information &
  T2002 &
  Web Interface &
  Use the Web Interface to gather mission documentation. \\
Reconnaissance &
  Gather Victim Org Information &
  T1591 &
  Web interface &
  Use the Web Interface to gather mission documentation. \\
Reconnaissance &
  In orbit proximity intelligence &
  T2029 &
  Threat Model Limitation &
  N.A. \\
Reconnaissance &
  Passive Interception (RF/Optical) &
  T2004 &
  Threat Model Limitation &
  N.A. \\
Reconnaissance &
  Phishing for Information &
  T1598 &
  Tangential &
  N.A. \\
Resource Development &
  Acquire or Build Infrastructure &
  T1583 &
  Telnet interface &
  Acquire ground segment using the Telnet service. \\
Resource Development &
  Compromise Account &
  T2038 &
  Tangential &
  N.A. \\
Resource Development &
  Compromise Infrastructure &
  T1584 &
  Threat Model Limitation &
  N.A. \\
Resource Development &
  Develop/Obtain Capabilities &
  T2007 &
  \begin{tabular}[c]{@{}l@{}}Ground software.\\ Flight software.\end{tabular} &
  \begin{tabular}[c]{@{}l@{}}Exploit OS, libraries or software vulnerabilities.\\ Deploy custom CSP application to forge TC/TM.\end{tabular} \\
Initial Access &
  Direct Attack to Space Communication Links &
  T2008 &
  \begin{tabular}[c]{@{}l@{}}Ground software.\\ Flight software.\end{tabular} &
  \begin{tabular}[c]{@{}l@{}}Use the ground software to send/receive TC/TM.\\ Deploy custom CSP application to forge TC/TM.\end{tabular} \\
Initial Access &
  Ground Segment Compromise &
  T2030 &
  \begin{tabular}[c]{@{}l@{}}Telnet interface.\\ Ground software.\end{tabular} &
  Use the telnet interface to access the ground software. \\
Initial Access &
  Supply Chain Compromise &
  T1195 &
  Tangential &
  N.A. \\
Initial Access &
  Trusted Relationship &
  T2039 &
  Threat Model Limitation &
  N.A. \\
Initial Access &
  Valid Credentials &
  T2009 &
  Tangential &
  N.A. \\
Execution &
  Modification of On Board Control Procedures modification &
  T2010 &
  \begin{tabular}[c]{@{}l@{}}Ground software.\\ Flight software.\end{tabular} &
  \begin{tabular}[c]{@{}l@{}}Upload a script to the satellite software:\\ tm\_send\_file code.py 1\\ 1: obc\_system python recv\_files/code.py\end{tabular} \\
Execution &
  Native API &
  T1106 &
  \begin{tabular}[c]{@{}l@{}}Ground software.\\ Flight software.\end{tabular} &
  \begin{tabular}[c]{@{}l@{}}Execute shell commands or delete system files:\\ 1: obc\_system \textless{}command\textgreater\\ 1: obc\_rm -r \$HOME\end{tabular} \\
Execution &
  Payload Exploitation to Execute Commands &
  T2012 &
  Tangential &
  N.A. \\
Persistence &
  Backdoor Installation &
  T2014 &
  \begin{tabular}[c]{@{}l@{}}Ground software.\\ Flight software.\end{tabular} &
  \begin{tabular}[c]{@{}l@{}}Upload a script to the satellite software:\\ tm\_send\_file backdoor.sh 1\\ 1: obc\_system ./recv\_files/backdoor.sh\end{tabular} \\
Persistence &
  Key Management Infrastructure Manipulation &
  T2013 &
  Tangential &
  N.A. \\
Persistence &
  Pre-OS Boot &
  T1542 &
  \begin{tabular}[c]{@{}l@{}}Ground software.\\ Flight software.\end{tabular} &
  \begin{tabular}[c]{@{}l@{}}Use obc\_system, obc\_rm, obc\_mkdir commands.\\ Upload/modify an OS configuration file.\\ Start/stop/schedule execution of services/daemons.\end{tabular} \\
Persistence &
  Valid Credentials &
  T2009 &
  Tangential &
  N.A. \\
Privilege Escalation &
  Become Avionics Bus Master &
  T2031 &
  Implementation limitation &
  N.A. \\
Privilege Escalation &
  Escape to Host &
  T1611 &
  \begin{tabular}[c]{@{}l@{}}Docker.\\ Virtual machine.\end{tabular} &
  Escape the container/VM with previously crafted exploits. \\
Defense Evasion &
  Impair Defenses &
  T1562 &
  \begin{tabular}[c]{@{}l@{}}Ground software.\\ Flight software.\end{tabular} &
  \begin{tabular}[c]{@{}l@{}}Send commands to change operation mode.\\ 1: drp\_set\_var\_name obc\_opmode 0\end{tabular} \\
Defense Evasion &
  Indicator Removal on Host &
  T1070 &
  \begin{tabular}[c]{@{}l@{}}Ground software.\\ Flight software.\end{tabular} &
  \begin{tabular}[c]{@{}l@{}}Use commands to remove artifacts, logs, etc.:\\ 1: obc\_rm \textless{}path\textgreater{}\end{tabular} \\
Defense Evasion &
  Masquerading &
  T2040 &
  \begin{tabular}[c]{@{}l@{}}Ground software.\\ Flight software.\end{tabular} &
  \begin{tabular}[c]{@{}l@{}}Use commands to upload artifacts modify system settings:\\ tm\_send\_file artifact 1\\ 1: obc\_mv artifact /etc/config/artifact\end{tabular} \\
Defense Evasion &
  Pre-OS Boot &
  T2041 &
  \begin{tabular}[c]{@{}l@{}}Ground software.\\ Flight software.\end{tabular} &
  \begin{tabular}[c]{@{}l@{}}Use obc\_system, obc\_rm, obc\_mkdir commands.\\ Upload/modify an OS configuration file.\\ Start/stop/schedule execution of services/daemons.\end{tabular} \\
Credential Access &
  Adversary in the Middle &
  T2042 &
  \begin{tabular}[c]{@{}l@{}}Ground software.\\ Flight software.\end{tabular} &
  Deploy a CSP application to capture/inject CSP packets. \\
Credential Access &
  Brute Force &
  T2043 &
  \begin{tabular}[c]{@{}l@{}}Ground software.\\ Flight software.\end{tabular} &
  \begin{tabular}[c]{@{}l@{}}Brute force valid TC parameters:\\ 1: obc\_ebf \textless{}KEY\textgreater{}\end{tabular} \\
Credential Access &
  Communication Link Sniffing &
  T2044 &
  Ground software. &
  \begin{tabular}[c]{@{}l@{}}Escape the ground software or docker and run tcpdump.\\ Deploy a CSP application to capture/inject CSP packets.\end{tabular} \\
Credential Access &
  Retrieve TT\&C master/session keys &
  T2015 &
  Tangential &
  N.A. \\
Discovery &
  Key Management Policy Discovery &
  T2032 &
  Tangential &
  N.A. \\
Discovery &
  Spacecraft's Components Discovery &
  T2034 &
  \begin{tabular}[c]{@{}l@{}}Ground software.\\ Flight software.\end{tabular} &
  \begin{tabular}[c]{@{}l@{}}Send TC to redirect satellite logs to ground segment:\\ 1: log\_set 5 2 10\end{tabular} \\
Discovery &
  System Service Discovery &
  T1007 &
  \begin{tabular}[c]{@{}l@{}}Ground software.\\ Flight software.\end{tabular} &
  \begin{tabular}[c]{@{}l@{}}Capture running processes information\\ 1: obc\_system ps -aux \textgreater ps.log\\ 1: tm\_send\_file 10 ps.log\end{tabular} \\
Discovery &
  Trust Relationships Discovery &
  T2033 &
  Tangential &
  N.A. \\
Lateral Movement &
  Compromise a Payload after compromising the main satellite platform &
  T2045 &
  Implementation Limitation &
  N.A. \\
Lateral Movement &
  \begin{tabular}[c]{@{}l@{}}  Compromise of satellite hypervisors \end{tabular} &
  T2017 &
  \begin{tabular}[c]{@{}l@{}}Docker.\\ Virtual machine.\end{tabular} &
  Escape the container/VM with previously crafted exploits. \\
Lateral Movement &
  Compromise the satellite platform starting from a compromised payload. &
  T2046 &
  Implementation Limitation &
  N.A. \\
Lateral Movement &
  Lateral Movement via common Avionics Bus. &
  T2016 &
  Implementation Limitation &
  N.A. \\
Collection &
  Adversary in the Middle &
  T1557 &
  \begin{tabular}[c]{@{}l@{}}Ground software.\\ Flight software.\end{tabular} &
  Deploy a CSP application to capture/inject CSP packets. \\
Collection &
  Data from link eavesdropping &
  T2018 &
  \begin{tabular}[c]{@{}l@{}}Ground software.\\ Flight software.\end{tabular} &
  \begin{tabular}[c]{@{}l@{}}Escape the ground software or docker and run tcpdump.\\ Deploy a CSP application to capture/inject CSP packets.\end{tabular} \\
Command and Control &
  Protocol Tunnelling &
  T2047 &
  \begin{tabular}[c]{@{}l@{}}Ground software.\\ Flight software.\end{tabular} &
  Deploy a malicious application that sends data over CSP \\
Command and Control &
  Telecommand a Spacecraft &
  T2019 &
  \begin{tabular}[c]{@{}l@{}}Ground software.\\ Flight software.\end{tabular} &
  \begin{tabular}[c]{@{}l@{}}Use the ground software to send TC:\\ 1: com\_ping 1\end{tabular} \\
Command and Control &
  TT\&C over ISL &
  T2048 &
  Threat Model Limitation &
  N.A. \\
Exfiltration &
  Exfiltration Over Payload Channel &
  T2021 &
  Implementation Limitation &
  N.A. \\
Exfiltration &
  Exfiltration Over TM Channel &
  T2022 &
  \begin{tabular}[c]{@{}l@{}}Ground software.\\ Flight software.\end{tabular} &
  The attacker deploys a custom CSP node or a backdoor \\
Exfiltration &
  Optical link modification &
  T2037 &
  Threat Model Limitation &
  N.A. \\
Exfiltration &
  RF modification &
  T2036 &
  Threat Model Limitation &
  N.A. \\
Exfiltration &
  Side-channel exfiltration &
  T2035 &
  Threat Model Limitation &
  N.A. \\
Impact &
  Data Manipulation &
  T2054 &
  \begin{tabular}[c]{@{}l@{}}Ground software.\\ Flight software.\end{tabular} &
  \begin{tabular}[c]{@{}l@{}}Send TC to modify/reset TM database:\\ 1: drp\_set\_var\_name drp\_ack\_ads 10000000\\ 1: drp\_reset\_payload 1 1010\\ 1: drp\_reset\_status 1010\end{tabular} \\
Impact &
  Ground Segment Jamming &
  T2050 &
  Threat Model Limitation &
  N.A. \\
Impact &
  Loss of spacecraft telecommanding &
  T2055 &
  \begin{tabular}[c]{@{}l@{}}Ground software.\\ Flight software.\end{tabular} &
  \begin{tabular}[c]{@{}l@{}}Send TC to change communication parameters.\\ Modify network configuration in the ground station.\end{tabular} \\
Impact &
  Permanent loss to telecommand satellite &
  T2027 &
  \begin{tabular}[c]{@{}l@{}}Ground software.\\ Flight software.\end{tabular} &
  \begin{tabular}[c]{@{}l@{}}Send TC to destroy filesystem:\\ 1: obc\_system rm -rf --no-preserve-root /\end{tabular} \\
Impact &
  Resource damage &
  T2028 &
  Threat Model Limitation &
  N.A. \\
Impact &
  Resource Hijacking &
  T1496 &
  \begin{tabular}[c]{@{}l@{}}Ground software.\\ Flight software.\end{tabular} &
  \begin{tabular}[c]{@{}l@{}}Upload a script to the satellite software:\\ tm\_send\_file code.py 1\\ 1: obc\_system python recv\_files/code.py\end{tabular} \\
Impact &
  Saturation of Inter Satellite Links &
  T2052 &
  Threat Model Limitation &
  N.A. \\
Impact &
  Saturation/Exhaustion of Spacecraft Resources &
  T2053 &
  \begin{tabular}[c]{@{}l@{}}Ground software.\\ Flight software.\end{tabular} &
  \begin{tabular}[c]{@{}l@{}}Send TC to create a reset loop:\\ 1: fp\_set\_cmd\_dt 10 2147483647 10 obc\_reset\end{tabular} \\
Impact &
  Service Stop &
  T1489 &
  \begin{tabular}[c]{@{}l@{}}Ground software.\\ Flight software.\end{tabular} &
  \begin{tabular}[c]{@{}l@{}}Send TC to launch a fork bomb or a reset loop\\ 1: obc\_system :()\{ :|:\& \};:\\ 1: fp\_set\_cmd\_dt 10 2147483647 10 obc\_reset\end{tabular} \\
Impact &
  Spacecraft Jamming &
  T2049 &
  Threat Model Limitation &
  N.A. \\
Impact &
  Temporary loss to telecommand satellite &
  T2026 &
  \begin{tabular}[c]{@{}l@{}}Ground software.\\ Flight software.\end{tabular} &
  \begin{tabular}[c]{@{}l@{}}Send TC to make the system unresponsive\\ 1: obc\_system sleep 3600\end{tabular} \\
Impact &
  Transmitted Data Manipulation &
  T2024 &
  Threat Model Limitation &
  N.A. \\ \bottomrule
\end{tabular}%
}
\end{table*}

%% file: 13-ae-appendix.tex
\lstset{
  backgroundcolor=\color{codegray},
  basicstyle=\ttfamily\footnotesize,
  commentstyle=\color{commentgreen}\itshape,
  emph={python,python3,docker,compose},       
  emphstyle=\color{purple}\bfseries,  
  breaklines=true,
  numberstyle=\tiny\color{gray},
  captionpos=b,
  numbers=none,
}

\section{Artifact Appendix}

This appendix accompanies the paper \textit{``HoneySat: A Network-based Satellite Honeypot Framework''} and provides detailed instructions for obtaining, installing, and evaluating the artifact submitted for NDSS 2026 Artifact Evaluation.
\subsection{Description \& Requirements}
Our artifact contains a Dockerized version of our honeypot framework, along with all the necessary components to bootstrap honeypots for satellite missions supporting two different space protocol stacks (CSP and CCSDS/YAMCS).
To aid in evaluations, we also included utility scripts that facilitate the evaluation of data provided by the honeypot.
\subsubsection{How to access}
The artifact can be found at \url{https://doi.org/10.5281/zenodo.17871431}.

\subsubsection{Hardware dependencies}
\begin{itemize}
    \item 25GB of disk space
    \item at least 4 CPU cores
    \item 8GB RAM
\end{itemize}

\subsubsection{Software dependencies}
Most Linux distributions will work. We verified the artifact on Ubuntu 24.04.2 LTS.
\begin{itemize}
    \item Docker (with compose and related components, we recommend to follow \url{https://docs.docker.com/engine/install/})
    \item Python 3.12 (more recent versions may also be compatible)
    \item Python virtual environment
    \item Bash
    \item Telnet client
    \item Any modern web browser
\end{itemize}

\subsubsection{Benchmarks} No External Benchmarks required.


\subsection{Artifact Installation \& Configuration}

\paragraph{Extract files}
Extract files from tar file downloaded from link:
\begin{lstlisting}
tar xzpvf ndss-artifact-eval.tar.gz
\end{lstlisting}

\paragraph{Repository Layout}
The root directory contains two subdirectories:
\begin{itemize}[leftmargin=*]
    \item \textbf{\texttt{deployment/}}: Dockerized HoneySat services with detailed setup documentation.
    \item \textbf{\texttt{evaluation/}}: Python utilities and scripts to help evaluate HoneySat's capabilities.
\end{itemize}
The evaluation utilities require Python~3.12 and the packages listed in \texttt{evaluation/requirements.txt}.

\paragraph{ Create and Activate a Python Virtual Environment.}
From the repository root:
\begin{lstlisting}
python3 -m venv .
source bin/activate
\end{lstlisting}

\paragraph{Install Python Dependencies}
Install the required packages for the evaluation utilities:
\begin{lstlisting}
python3 -m pip install -r evaluation/requirements.txt
\end{lstlisting}

\paragraph{Configure Docker user}
We need to add the current user to the Docker group to avoid using sudo with Docker all the time.
\begin{lstlisting}
sudo usermod -aG docker $USER
sudo reboot now
\end{lstlisting}

\paragraph{Start the Honeypot Services.}
All remaining setup is handled by Docker containers. You can either:
\begin{itemize}[leftmargin=*]
    \item Run the provided convenience scripts (see ``Quick Start for Pass Simulations'' below), \emph{or}
    \item Follow the guide in \textbf{\texttt{deployment/README.md}} to bring up the services via Docker Compose.
\end{itemize}

\paragraph{Quick Start for Pass Simulations}
A key HoneySat feature is its realistic communication windows: the satellite is reachable only while passing over a ground station. Many experiments benefit from a setup in which the satellite becomes reachable immediately or within a short time window. We provide two helper scripts that compute a suitable ground-station location along the satellite's predicted ground track so that a pass occurs shortly after startup. Each script builds and starts the necessary Docker Compose services and will shut them down cleanly on \texttt{CTRL+C}.

\begin{itemize}[leftmargin=*]
    \item \textbf{CSP-based honeypot:}
\begin{lstlisting}
./evaluation/experiment-1/run-experiment-csp.sh
\end{lstlisting}
     This script builds and starts a CSP honeypot instance,  and positions the ground station to enable communication almost immediately, allowing you to observe how an attacker would perceive a pass without waiting for a real one.

    \item \textbf{CCSDS/YAMCS-based honeypot:}
\begin{lstlisting}
./evaluation/experiment-1/run-experiment-ccsds.sh
\end{lstlisting}
    This script builds and starts a CCSDS and YAMCS-based honeypot instance, using the same predicted ground-station placement to trigger an imminent pass.
\end{itemize}

\noindent Please proceed with the experiments described in the following sections, or consult \textbf{\texttt{deployment/README.md}} for additional configuration details.

\subsection{Experiment Workflow}
The following experiments all involve starting the honeypot system and interacting with it in some way (either programmatically or manually). The workflow typically involves issuing a startup command and then performing tasks outlined in the experiment description. 
For some tasks, timing matters; for example, the TCs have to be issued while the simulated satellite is reachable via the simulated ground station.


\subsection{Major Claims}
We make the following claims in our paper
\begin{itemize}
    \item (C1): \textsc{HoneySat} can be used to deceive adversaries and log activities.
    \begin{itemize}
        \item HoneySat's simulator provides believable TM (Experiment 1.1)
        \item HoneySat has realistic communication windows (Experiment 1.2)
        \item HoneySat provides interaction capabilities (process TC, provide TM) (experiment 1.3)
        \item HoneySat Logs interaction details (Experiment 1.4)
    \end{itemize}
    \item (C2): \textsc{HoneySat} Is extensible and supports two different protocol ecosystems
    \begin{itemize}
        \item HoneySat is configurable (Experiment 2)
    \end{itemize}
\end{itemize}

\subsection{Evaluation}

\subsubsection{Experiment 1 (E1.1)}
[Believable TM/TC] \; [10 human-minutes]

\textbf{Goal.} Demonstrate that HoneySat produces believable telemetry (TM) in response to telecommands (TC) for a CCSDS/YAMCS setup.

\textbf{Preparation.} Ensure dependencies are installed as listed above.

\textbf{Execution.}
\begin{lstlisting}
./evaluation/experiment-1/run-experiment-ccsds.sh
\end{lstlisting}
After the script completes and prints telemetry values, you may also inspect the telemetry via the YAMCS web interface.

\textbf{Expected Results.} Believable current, voltage, and temperature values for the selected battery configuration:
\begin{itemize}[leftmargin=*]
  \item Voltage: ~8000 mV (normal test case)
  \item Temperature: ~30 °C
  \item Current draw: ~74 mA
\end{itemize}

\vspace{0.6em}
\subsubsection{Experiment 1.2 (E1.2)}
[Believable passes] \; [20 human-minutes]

\textbf{Goal.} Show that HoneySat enforces realistic communication windows: the satellite is reachable only during predicted passes over a ground station.

\textbf{Preparation.}
\begin{lstlisting}
./evaluation/experiment-1/run-experiment-csp.sh
\end{lstlisting}
This positions the ground station so that a pass begins approximately 2-3 minutes after startup (adjustable), and starts all required services.

\textbf{Execution.}
\begin{enumerate}[leftmargin=*]
  \item Open the web interface at \url{http://localhost:80}. Log in with username \texttt{admin} and password \texttt{admin}. Click the ground-station icon and view the next predicted passes. The pass should soon show as ``ongoing.''
  \item In a separate terminal, connect to the CSP telnet interface and activate it:
\begin{lstlisting}
telnet localhost 24
# In the telnet session:
activate
\end{lstlisting}
  \item Probe satellite reachability every few seconds:
\begin{lstlisting}
1: com_ping 10
\end{lstlisting}
\end{enumerate}

\textbf{Expected Results.} The satellite responds to pings only during the predicted pass and not before or after. Responses will indicate the expected addressing (source address 1, destination address 10).

\vspace{0.6em}
\subsubsection{Experiment 1.3 (E1.3)}
[Simulate Interaction] \; [15 human-minutes]

\textbf{Goal.} Demonstrate interactive capabilities once the satellite becomes reachable.

\textbf{Preparation.} Repeat the setup from Experiment~1.2. Review the command reference in the \emph{experiment-1.3} directory.

\textbf{Execution.} After the satellite becomes reachable (as verified in Experiment~1.2), issue commands via the telnet interface. For example, to execute arbitrary shell commands on the OBC:
\begin{lstlisting}
1: obc_system [shell command]
\end{lstlisting}

\textbf{Expected Results.} The experimenter can successfully execute arbitrary shell commands via \texttt{obc\_system} while the satellite is reachable.

\vspace{0.6em}
\subsubsection{Experiment 1.4 (E1.4)}
[View logging capabilities] \; [5 human-minutes]

\textbf{Goal.} Verify that HoneySat logs interactions and relevant parameters.

\textbf{Preparation.} Keep a HoneySat CSP instance running (e.g., from Experiment~1.3), optionally after interacting with it.

\textbf{Execution.}
\begin{lstlisting}
python3 ./evaluation/experiment-1/python_dump_mongodb/dump_mongodb.py
\end{lstlisting}
Alternatively, connect to the MongoDB instance with a database client of your choice.

\textbf{Expected Results.} The script prints MongoDB contents to \texttt{stdout}, showing logged parameters and interaction details.

\vspace{0.6em}
\subsubsection{Experiment 2 (E2)}
[Customization]

\textbf{Goal.} Demonstrate HoneySat’s configurability across protocol stacks and scenarios.

\textbf{Preparation.} Navigate to \emph{./evaluation/experiment-2}.

\textbf{Execution.}
\begin{enumerate}[leftmargin=*]
  \item Create a baseline customization:
\begin{lstlisting}
python3 ./honeysat.py [csp|ccsds] "{satellite name}" "{location}"
\end{lstlisting}
For example:
\begin{lstlisting}
python3 ./honeysat.py csp "BEESAT" "Berlin"
\end{lstlisting}
  \item Start services with the generated configuration (or configs in this directory):
\begin{lstlisting}
python3 ./honeysat.py start [csp|ccsds] "{satellite name}" "{location}"
\end{lstlisting}
For example:
\begin{lstlisting}
python3 ./honeysat.py start csp "BEESAT" "Berlin"
\end{lstlisting}
  \item When finished, stop the services:
\begin{lstlisting}
python3 ./honeysat.py stop [csp|ccsds]
\end{lstlisting}
For example:
\begin{lstlisting}
python3 ./honeysat.py stop csp
\end{lstlisting}
\end{enumerate}

\textbf{Expected Results.} HoneySat can be configured with relatively low time effort to reflect different satellites, locations, and protocol ecosystems.

%% file: bib.bib
@misc{estcube2014telemetry,
  author       = {{ESTCube-1 Team}},
  title        = {Telemetry Packet Description},
  year         = {2014},
  howpublished = {\url{https://web.archive.org/web/20140808061141/http://www.estcube.eu/en/telemetry-packet-description}},
  note         = {Accessed: 2025-07-22},
}

@book{spitzner2002honeypots,
  title={Honeypots: tracking hackers},
  author={Spitzner, Lance},
  year={2002},
  publisher={Addison-Wesley Longman Publishing Co., Inc.}
}

@techreport{arc2018opcinstalledbase,
  title        = {OPC Installed Base Insights},
  author       = {{ARC Advisory Group}},
  institution  = {OPC Foundation},
  year         = {2018},
  url          = {https://opcfoundation.org/wp-content/uploads/2018/02/ARC-Report-OPC-Installed-Base-Insights.pdf},
  note         = {Accessed October 2025}
}

@inproceedings{ma2023network,
  title={Network characteristics of leo satellite constellations: A starlink-based measurement from end users},
  author={Ma, Sami and Chou, Yi Ching and Zhao, Haoyuan and Chen, Long and Ma, Xiaoqiang and Liu, Jiangchuan},
  booktitle={IEEE INFOCOM 2023-IEEE Conference on Computer Communications},
  pages={1--10},
  year={2023},
  organization={IEEE}
}

@techreport{uscc2011annualreport,
  title        = {2011 Annual Report to Congress},
  author       = {{U.S.-China Economic and Security Review Commission}},
  institution  = {U.S. Government Printing Office},
  year         = {2011},
  url          = {https://www.uscc.gov/sites/default/files/annual_reports/annual_report_full_11.pdf},
  note         = {Accessed October 2025}
}

@online{livescience2024satellites,
  title        = {How Many Satellites Could Fit in Earth Orbit---and How Many Do We Really Need?},
  author       = {Wall, Mike},
  year         = {2024},
  month        = {July},
  url          = {https://www.livescience.com/space/space-exploration/how-many-satellites-could-fit-in-earth-orbit-and-how-many-do-we},
  note         = {Accessed October 2025, Live Science}
}

@inproceedings {wireless2024bisping,
author = {Robin Bisping and Johannes Willbold and Martin Strohmeier and Vincent Lenders},
title = {Wireless Signal Injection Attacks on {VSAT} Satellite Modems},
booktitle = {33rd USENIX Security Symposium (USENIX Security 24)},
year = {2024},
isbn = {978-1-939133-44-1},
address = {Philadelphia, PA},
pages = {6075--6091},
url = {https://www.usenix.org/conference/usenixsecurity24/presentation/bisping},
publisher = {USENIX Association},
month = aug
}

@misc{eoportal2025pixl1,
        author = {{eoPortal}},	
        title = {PIXL-1 / Formerly CubeL or OSIRIS4CubeSat
},
	copyright = {},
	url = {https://www.eoportal.org/satellite-missions/pixl-1},
	urldate = {2025-01-07},
	publisher = {},
	month = Jan,
	year = {2025},
}

@misc{nasa2025acs3,
        author = {{NASA}},	
        title = {Advanced Composite Solar Sail System (ACS3)},
	copyright = {},
	url = {https://www.nasa.gov/mission/acs3/},
	urldate = {2025-01-07},
	publisher = {},
	month = Jan,
	year = {2025},
}

@misc{likert2025qualtrics,
        author = {{Qualtrics}},	
        title = {What is a likert scale?},
	copyright = {},
	url = {https://www.qualtrics.com/experience-management/research/likert-scale/},
	urldate = {2025-01-07},
	publisher = {},
	month = Jan,
	year = {2025},
}

@misc{esa2023oversees,
        author = {{The European Space Agency}},	
        title = {ESA oversees in-orbit cybersecurity demonstration},
	copyright = {},
	url = {https://www.esa.int/Enabling_Support/Operations/ESA_oversees_in-orbit_cybersecurity_demonstration},
	urldate = {2024-12-20},
	publisher = {},
	month = May,
	year = {2023},
}

@misc{ossman2024tigervnc,
        author = {Ossman, Pierre},	
        title = {TigerVNC/tigervnc: High performance, multi-platform VNC client and server},
	copyright = {},
	url = {https://github.com/TigerVNC/tigervnc},
	urldate = {2024-09-07},
	publisher = {},
	month = Jul,
	year = {2024},
}

@misc{merri2007cutting,
        author = {Merri, Mario and Ercolani, Alessandro and Guerrucci, Damiano and Reggestad, Vemund and Verrier, David},	
        title = {Cutting the Cost of ESA Mission Ground Software},
	copyright = {},
	url = {https://www.esa.int/esapub/bulletin/bulletin130/bul130g_merri.pdf},
	urldate = {2024-07-7},
	publisher = {},
	month = May,
	year = {2007},
}

@misc{csete2023gpredict,
        author = {Csete, Alexandru},	
        title = {Gpredict: Free, Real-Time Satellite Tracking and Orbit Prediction Software},
	copyright = {},
	url = {https://oz9aec.dk/gpredict/},
	urldate = {2024-07-7},
	publisher = {},
	month = Dec,
	year = {2023},
}

@misc{nasa2025smallsats,
        author = {{NASA}},	
        title = {What are SmallSats and CubeSats?},
	copyright = {},
	url = {https://www.nasa.gov/what-are-smallsats-and-cubesats/},
	urldate = {2025-01-17},
	publisher = {NASA}
}

@misc{nasa2024flight,
        author = {{NASA Goddard Space Flight Center}},	
        title = {Flight Training: Introduction},
	copyright = {},
	url = {https://solc.gsfc.nasa.gov/modules/missionops/mainMenu_textOnly.php},
	urldate = {2024-06-24},
	publisher = {}
}

@misc{california2024earth,
        author = {California Polytechnic State University, San Luis Obispo},	
        title = {Earth Station - PolySat},
	copyright = {},
	url = {https://www.polysat.org/earth-station},
	urldate = {2024-06-24},
	publisher = {},
	month = Apr,
	year = {2024},
}

@misc{mitre2024groups,
        author = {{The MITRE Corporation}},	
        title = {Groups | MITRE ATT\&CK},
	copyright = {},
	url = {https://attack.mitre.org/groups/},
	urldate = {2024-04-28},
	publisher = {The MITRE Corporation},
	month = Apr,
	year = {2024},
}

@inproceedings{scharnowski2023case,
  title={A Case Study on Fuzzing Satellite Firmware},
  author={Scharnowski, Tobias and Buchmann, Felix and W{\"o}rner, Simon and Holz, Thorsten},
  booktitle={Workshop on the Security of Space and Satellite Systems (SpaceSec)},
  year={2023}
}

@misc{amazon2024satellite,
        author = {{Amazon Web Services}},	
        title = {Satellite As A Service  - AWS Ground Station - AWS},
	copyright = {},
	url = {https://aws.amazon.com/ground-station/},
	urldate = {2024-04-19},
	publisher = {Amazon Web Services},
	month = Mar,
	year = {2024},
}

@misc{semanik2023private,
        author = {Semanik, Mitch and Crotty, Patrick},
        title = {U.S. Private Space Launch Industry is Out of this World},
	copyright = {},
	url = {https://www.usitc.gov/publications/332/executive_briefings/ebot_us_private_space_launch_industry_is_out_of_this_world.pdf},
	urldate = {2024-04-24},
	publisher = {},
	month = Nov,
	year = {2023},
}

@inproceedings{ivancic2003architecture,
  title={Architecture and system engineering development study of space-based satellite networks for nasa missions},
  author={Ivancic, William D},
  booktitle={2003 Aerospace Conference},
  number={NASA/TM-2003-212187},
  year={2003}
}

@misc{holmes2024growing,
        author = {Holmes, Mark},
        title = {The Growing Risk of a Major Satellite Cyber Attack},
	copyright = {},
	url = {https://interactive.satellitetoday.com/the-growing-risk-of-a-major-satellite-cyber-attack/},
	urldate = {2024-04-24},
	publisher = {},
	month = May,
	year = {2023},
}

@misc{geological2022earthexplorer,
        author = {{U.S. Geological Survey}},
        title = {EarthExplorer | U.S. Geological Survey},
	copyright = {},
	url = {https://www.usgs.gov/tools/earthexplorer},
	urldate = {2024-04-23},
	publisher = {University of Southern California},
	month = Nov,
	year = {2022},
}

@misc{space2023global,
        author = {{United States Space Force}},	
        title = {Global Positioning System > Space Operations Command (SpOC) > Display},
	copyright = {},
	url = {https://www.spoc.spaceforce.mil/About-Us/Fact-Sheets/Display/Article/2381726/global-positioning-system},
	urldate = {2024-04-21},
	publisher = {Official United States Space Force Website},
	month = Feb,
	year = {2023},
}

@INPROCEEDINGS{daubert2018honeydrone,
  author={Daubert, Jorg and Boopalan, Dhanasekar and Mühlhäuser, Max and Vasilomanolakis, Emmanouil},
  booktitle={NOMS 2018 - 2018 IEEE/IFIP Network Operations and Management Symposium}, 
  title={HoneyDrone: A medium-interaction unmanned aerial vehicle honeypot}, 
  year={2018},
  volume={},
  number={},
  pages={1-6},
  doi={10.1109/NOMS.2018.8406315}}

@INPROCEEDINGS{conti2022icspot,
  author={Conti, Mauro and Trolese, Francesco and Turrin, Federico},
  booktitle={2022 International Symposium on Networks, Computers and Communications (ISNCC)}, 
  title={ICSpot: A High-Interaction Honeypot for Industrial Control Systems}, 
  year={2022},
  volume={},
  number={},
  pages={1-4},
  doi={10.1109/ISNCC55209.2022.9851732}
}

@inproceedings{lucchese2023honeyics,
author = {Lucchese, Marco and Lupia, Francesco and Merro, Massimo and Paci, Federica and Zannone, Nicola and Furfaro, Angelo},
title = {HoneyICS: A High-interaction Physics-aware Honeynet for Industrial Control Systems},
year = {2023},
isbn = {9798400707728},
publisher = {Association for Computing Machinery},
address = {New York, NY, USA},
url = {https://doi.org/10.1145/3600160.3604984},
doi = {10.1145/3600160.3604984},
booktitle = {Proceedings of the 18th International Conference on Availability, Reliability and Security},
articleno = {113},
numpages = {10},
series = {ARES '23}
}

@misc{nazario2024awesome,
        author = {Nazario, Jose},	
        title = {paralax/awesome-honeypots: an awesome list of honeypot resources},
	copyright = {},
	url = {https://github.com/paralax/awesome-honeypots},
	urldate = {2024-04-21},
	publisher = {GitHub},
	month = Mar,
	year = {2024},
}

@article{ilg2023survey,
title = {A survey of contemporary open-source honeypots, frameworks, and tools},
journal = {Journal of Network and Computer Applications},
volume = {220},
pages = {103737},
year = {2023},
issn = {1084-8045},
doi = {https://doi.org/10.1016/j.jnca.2023.103737},
url = {https://www.sciencedirect.com/science/article/pii/S108480452300156X},
author = {Niclas Ilg and Paul Duplys and Dominik Sisejkovic and Michael Menth},
}

@inproceedings{icsnet2024salazar,
author = {Salazar, Luis and L\'{o}pez-Morales, Efr\'{e}n and Lozano, Juan and Rubio-Medrano, Carlos and C\'{a}rdenas, \'{A}lvaro A.},
title = {ICSNet: A Hybrid-Interaction Honeynet for Industrial Control Systems},
year = {2024},
isbn = {9798400712449},
publisher = {Association for Computing Machinery},
address = {New York, NY, USA},
url = {https://doi.org/10.1145/3690134.3694813},
doi = {10.1145/3690134.3694813},
booktitle = {Proceedings of the Sixth Workshop on CPS\&IoT Security and Privacy},
pages = {68–79},
numpages = {12},
location = {Salt Lake City, UT, USA},
series = {CPSIoTSec'24}
}

@misc{nmap2025nmap,
        author = {Nmap},	
        title = {OS Detection},
	copyright = {},
	url = {https://nmap.org/book/man-os-detection.html},
	urldate = {2025-01-18},
	publisher = {Nmap},
	month = Jan,
	year = {2025},
}

@misc{oosterhof2024cowrie,
        author = {Oosterhof, Michel},	
        title = {cowrie/cowrie: Cowrie SSH/Telnet Honeypot https://cowrie.readthedocs.io},
	copyright = {},
	url = {https://github.com/cowrie/cowrie},
	urldate = {2024-04-21},
	publisher = {GitHub},
	month = Apr,
	year = {2024},
}

@inproceedings{lopez2022honeyplc,
author = {L\'{o}pez-Morales, Efr\'{e}n and Rubio-Medrano, Carlos and Doup\'{e}, Adam and Shoshitaishvili, Yan and Wang, Ruoyu and Bao, Tiffany and Ahn, Gail-Joon},
title = {HoneyPLC: A Next-Generation Honeypot for Industrial Control Systems},
year = {2020},
isbn = {9781450370899},
publisher = {Association for Computing Machinery},
address = {New York, NY, USA},
url = {https://doi.org/10.1145/3372297.3423356},
doi = {10.1145/3372297.3423356},
booktitle = {Proceedings of the 2020 ACM SIGSAC Conference on Computer and Communications Security},
pages = {279–291},
numpages = {13},
location = {Virtual Event, USA},
series = {CCS '20}
}

@misc{vestegaard2014conpot,
        author = {Vestergaard, Johnny},	
        title = {mushorg/conpot: ICS/SCADA honeypot},
	copyright = {},
	url = {https://github.com/mushorg/conpot},
	urldate = {2024-04-21},
	publisher = {GitHub},
	month = Mar,
	year = {2024},
}

@misc{esa2013telemetry,
        author = {{The European Space Agency}},	
        title = {ESA - Telemetry \& Telecommand},
	copyright = {},
	url = {https://www.esa.int/Enabling_Support/Space_Engineering_Technology/Onboard_Computers_and_Data_Handling/Telemetry_Telecommand},
	urldate = {2024-04-21},
	publisher = {},
	month = Mar,
	year = {2013},
}

@INPROCEEDINGS{acharya2024conning,
        author = {B. Acharya and M. Saad and A. Emanuele Cinà and L. Schönherr and H. Dai Nguyen and A. Oest and P. Vadrevu and T. Holz},
        booktitle = {2024 IEEE Symposium on Security and Privacy (SP)},
        title = {Conning the Crypto Conman: End-to-End Analysis of Cryptocurrency-based Technical Support Scams},
        year = {2024},
        doi = {10.1109/SP54263.2024.00156},
        url = {https://doi.ieeecomputersociety.org/10.1109/SP54263.2024.00156},
        publisher = {IEEE Computer Society},
        address = {Los Alamitos, CA, USA},
        month = {may}
}

@misc{mongodb2024mongodb,
        author = {{MongoDB, Inc.}},	
        title = {MongoDB: The Developer Data Platform | MongoDB},
	copyright = {© 2024 MongoDB, Inc.},
	url = {https://www.mongodb.com/},
	urldate = {2024-04-19},
	publisher = {MongoDB, Inc.},
	month = Apr,
	year = {2024},
}

@misc{wood2006introduction,
        author = {Wood, Lloyd},	
        title = {Introduction to satellite constellations: orbital types, uses and related facts},
	copyright = {},
	url = {https://savi.sourceforge.io/about/lloyd-wood-isu-summer-06-constellations-talk.pdf},
	urldate = {2024-04-18},
	publisher = {},
	month = Jul,
	year = {2006},
}

@misc{nasa2019remote,
        author = {{NASA}},	
        title = {What is Remote Sensing? | Earthdata},
	copyright = {},
	url = {https://www.earthdata.nasa.gov/learn/backgrounders/remote-sensing},
	urldate = {2024-04-18},
	publisher = {},
	month = Aug,
	year = {2019},
}

@misc{esa2018payload,
        author = {{The European Space Agency}},	
        title = {ESA - About Payload Systems},
	copyright = {},
	url = {https://www.esa.int/Enabling_Support/Space_Engineering_Technology/About_Payload_Systems},
	urldate = {2024-04-17},
	publisher = {},
	month = Apr,
	year = {2024},
}

@misc{yost2023nasa,
        author = {Yost, Bruce D.},	
        title = {NASA SSRI Knowledge Base | Detailed Design and Analysis > Subsystem Design > Command and Data Handling},
	copyright = {},
	url = {https://s3vi.ndc.nasa.gov/ssri-kb/topics/32/},
	urldate = {2024-04-17},
	publisher = {},
	month = Jun,
	year = {2023},
}

@article{vancamp2022world,
title = {A World without Satellite Data as a Result of a Global Cyber-Attack},
journal = {Space Policy},
volume = {59},
pages = {101458},
year = {2022},
issn = {0265-9646},
doi = {https://doi.org/10.1016/j.spacepol.2021.101458},
url = {https://www.sciencedirect.com/science/article/pii/S0265964621000503},
author = {Charlotte, {Van Camp} and Walter, Peeters},
}

@inproceedings{marra2024feasibility,
  title={On the Feasibility of CubeSats Application Sandboxing for Space Missions},
  author={Marra, Gabriele and Planta, Ulysse and W{\"u}stenberg, Philipp and Abbasi, Ali},
  booktitle={Workshop on the Security of Space and Satellite Systems (SpaceSec)},
  year={2024}
}

@misc{cohen1998deception,
        author = {Cohen, Frederick},	
        title = {Deception ToolKit},
	copyright = {},
	url = {http://all.net/dtk/},
	urldate = {2024-04-12},
	publisher = {},
	month = Mar,
	year = {1998},
}

@misc{stingar2024clever,
        author = {{Shared Threat Intelligence for Network Gatekeeping and Automated Response (STINGAR)}},	
        title = {About - STINGAR},
	copyright = {},
	url = {https://stingar.security.duke.edu/about-2/},
	urldate = {2024-04-12},
	publisher = {Duke University},
	month = Apr,
	year = {2024},
}

@article{hilt2020caught,
  title={Caught in the act: Running a realistic factory honeypot to capture real threats},
  author={Hilt, Stephen and Maggi, Federico and Perine, Charles and Remorin, Lord and R{\"o}sler, Martin and Vosseler, Rainer},
  journal={Trend Micro Research},
  year={2020}
}

@misc{burgess2023clever,
        author = {Burgess, Matt},	
        title = {A Clever Honeypot Tricked Hackers Into Revealing Their Secrets},
	copyright = {2024 Condé Nast. All rights reserved.},
	url = {https://www.wired.com/story/hacker-honeypot-go-secure/},
	urldate = {2024-04-12},
	publisher = {Wired},
	month = Aug,
	year = {2023},
}

@misc{franceschi2023thousands,
        author = {Franceschi-Bicchierai, Lorenzo},	
        title = {Thousands of new honeypots deployed across Israel to catch hackers},
	copyright = {2024 Yahoo.All rights reserved.},
	url = {https://techcrunch.com/2023/11/20/thousands-of-new-honeypots-deployed-across-israel-to-catch-hackers/},
	urldate = {2024-04-12},
	publisher = {TechCrunch},
	month = Nov,
	year = {2023},
}

@article{cohen2006use,
  title={The use of deception techniques: Honeypots and decoys},
  author={Cohen, Fred},
  journal={Handbook of Information Security},
  volume={3},
  number={1},
  pages={646--655},
  year={2006},
  publisher={John Wiley \& Sons Chichester}
}

@misc{granger2024pyzmq,
        author = {Granger, Brian E. and Ragan-Kelley, Min},	
        title = {PyZMQ Documentation},
	copyright = {Creative Commons Attribution-Share Alike 3.0 License},
	url = {https://pyzmq.readthedocs.io/en/latest/},
	urldate = {2024-04-11},
	publisher = {},
	month = apr,
	year = {2024},
}

@misc{docker2024docker,
        author = {{ Docker Inc.}},	
        title = {Docker: Accelerated Container Application Development},
	url = {https://www.docker.com/},
	urldate = {2024-04-11},
	publisher = {{ Docker Inc.}},
	month = apr,
	year = {2024},
}

@INPROCEEDINGS{willbold2023space,
  author={Willbold, Johannes and Schloegel, Moritz and Vögele, Manuel and Gerhardt, Maximilian and Holz, Thorsten and Abbasi, Ali},
  booktitle={2023 IEEE Symposium on Security and Privacy (SP)}, 
  title={Space Odyssey: An Experimental Software Security Analysis of Satellites}, 
  year={2023},
  volume={},
  number={},
  pages={1-19},
  doi={10.1109/SP46215.2023.10351029}
}

@misc{nist2024tactics,
        author = {{National Institute of Standards and Technology (NIST)}},	
        title = {tactics, techniques, and procedures (TTP) - Glossary},
	url = {https://csrc.nist.gov/glossary/term/tactics_techniques_and_procedures},
	urldate = {2024-04-10},
	publisher = {National Institute of Standards and Technology (NIST)},
	month = apr,
	year = {2024},
}

@misc{raza2024what,
        author = {Raza, Muhammad},	
        title = {What Are TTPs? Tactics, Techniques \& Procedures Explained},
	url = {https://www.splunk.com/en_us/blog/learn/ttp-tactics-techniques-procedures.html},
	urldate = {2024-04-10},
	publisher = {splunk},
	month = apr,
	year = {2024},
}

@inproceedings{vallado2012two,
  title={Two-line element sets--practice and use},
  author={Vallado, David A and Cefola, Paul J},
  booktitle={63rd International Astronautical Congress},
  pages={1--14},
  year={2012},
  publisher={International Astronautical Federation},
  address={Naples, Italy}
}

@misc{gomspace2021nanocom,
        author= {GomSpace},
	title = {NanoCom MS100 Datasheet},
	url = {https://gomspace.com/UserFiles/Subsystems/datasheet/gs-ds-nanocom-ms100-13.pdf},
	language = {en-US},
	urldate = {2024-03-25},
	month = mar,
	year = {2021},
}

@misc{esa2024space-shield,
        author = {European Space Agency},
        year = {2023},
	title = {{ESA} {SPACE}-{SHIELD}},
	url = {https://spaceshield.esa.int/#},
        publisher = {European Space Agency},
	urldate = {2024-03-15},
}

@misc{suchai-flight-software,
        author = {{Space and Planetary Exploration Laboratory, University of Chile}},
	title = {{SPEL} / {SUCHAI}-{Flight}-{Software} · {GitLab}},
	url = {https://gitlab.com/spel-uchile/suchai-flight-software},
	language = {en},
	urldate = {2024-03-14},
	journal = {GitLab},
	month = feb,
	year = {2024},
}

@inproceedings{mccomas2016core,
  title={The core flight system (cFS) community: Providing low cost solutions for small spacecraft},
  author={McComas, David and Wilmot, Jonathan and Cudmore, Alan},
  booktitle={Annual AIAA/USU Conference on Small Satellites},
  publisher = {{Utah State University}, {University Libraries}},
  year={2016},
  address={Logan, UT}
}

@ARTICLE{gonzalez2019architecture,
  author={Gonzalez, Carlos E. and Rojas, Camilo J. and Bergel, Alexandre and Diaz, Marcos A.},
  journal={IEEE Access}, 
  title={An Architecture-Tracking Approach to Evaluate a Modular and Extensible Flight Software for CubeSat Nanosatellites}, 
  year={2019},
  volume={7},
  number={},
  pages={126409-126429},
  doi={10.1109/ACCESS.2019.2927931}
}

@inproceedings{holz2005detecting,
  title={Detecting honeypots and other suspicious environments},
  author={Holz, Thorsten and Raynal, Frederic},
  booktitle={Proceedings from the sixth annual IEEE SMC information assurance workshop},
  pages={29--36},
  year={2005},
  publisher={IEEE},
  address = {West Point, NY, USA},
}

@article{pavur2022building,
  title={Building a launchpad for satellite cyber-security research: lessons from 60 years of spaceflight},
  author={Pavur, James and Martinovic, Ivan},
  journal={Journal of Cybersecurity},
  volume={8},
  number={1},
  pages={tyac008},
  year={2022},
  publisher={Oxford University Press}
}

@book{provos2007virtual,
  title={Virtual honeypots: from botnet tracking to intrusion detection},
  author={Provos, Niels and Holz, Thorsten},
  year={2007},
  address = {Boston, MA, USA},
  publisher={Pearson Education}
}

@misc{SpacePacketProtocol2020,
        author= {The Consultative Committee for Space Data Systems (CCSDS)},
	title = {Space {Packet} {Protocol}},
	language = {en},
	year = {2020},
	url = {https://public.ccsds.org/Pubs/133x0b2e1.pdf},
}

@misc{ProtocolStackCubesatZMQ,
        author = {Yasushi Shoji},
	title = {The {Protocol} {Stack} — {Cubesat} {Space} {Protocol}},
	url = {https://libcsp.github.io/libcsp/protocolstack.html},
	urldate = {2024-02-28},
        year = {2024}
}

@misc{LibcspLibcsp2024,
        author = {Yasushi Shoji},
	title = {libcsp/libcsp},
	copyright = {MIT},
	url = {https://github.com/libcsp/libcsp},
	urldate = {2024-02-28},
	publisher = {libcsp},
	month = feb,
	year = {2024},
}

@article{maya1_inquirer_2018,
  author       = {Salces, Adrian and Javier, Joven},
  title        = {Maya‑1: Cube satellite latest Pinoy venture into space},
  journal      = {Philippine Daily Inquirer},
  year         = {2018},
  month        = {Jul},
  day          = {01},
  howpublished = {\url{https://technology.inquirer.net/77081/maya-1-cube-satellite-latest-pinoy-venture-space}},
  note         = {Accessed: 2025‑08‑06}
}

@misc{uwe1_wuerzburg_2005,
  title        = {UWE‑1 (Universität Würzburg’s Experimentalsatellit‑1)},
  author       = {{Gunter’s Space Page}},
  howpublished = {\url{https://space.skyrocket.de/doc_sdat/uwe-1.htm}},
  note         = {Accessed: 2025‑08‑06},
  year         = {2005}
}

@misc{GetStarted,
        author = {{The ZeroMQ authors}},
	title = {{ZeroMQ | }Get started},
	url = {https://zeromq.org/get-started/},
	urldate = {2024-02-28},
        year = {2024},
}

@inproceedings{Garrido2023,
  title={The First Chilean Satellite Swarm: Approach and Lessons Learned},
  author={Garrido, Cristobal and Obreque, Elias and Vidal-Valladares, Matias and Gutierrez, Samuel and Diaz Quezada, Marcos},
  year={2023},
  booktitle={AIAA/USU Conference on Small Satellites, Year in Review - Research \& Academia, SSC23-WVII-07},
  address={Logan, UT},
  url={https://digitalcommons.usu.edu/smallsat/2023/all2023/56/}
}

@inproceedings {298052,
author = {Efr{\'e}n L{\'o}pez-Morales and Ulysse Planta and Carlos Rubio-Medrano and Ali Abbasi and Alvaro A. Cardenas},
title = {{SoK}: Security of Programmable Logic Controllers},
booktitle = {33rd USENIX Security Symposium (USENIX Security 24)},
year = {2024},
isbn = {978-1-939133-44-1},
address = {Philadelphia, PA},
pages = {7103--7122},
url = {https://www.usenix.org/conference/usenixsecurity24/presentation/lopez-morales},
publisher = {USENIX Association},
month = aug
}

@inproceedings{tay2023taxonomy,
  title={Taxonomy of fingerprinting techniques for evaluation of smart grid honeypot realism},
  author={Tay, Vanessa and Li, Xinran and Mashima, Daisuke and Ng, Bennet and Cao, Phuong and Kalbarczyk, Zbigniew and Iyer, Ravishankar K},
  booktitle={2023 IEEE International Conference on Communications, Control, and Computing Technologies for Smart Grids (SmartGridComm)},
  pages={1--7},
  year={2023},
  organization={IEEE}
}

@inproceedings{kyung2017honeyproxy,
  title={HoneyProxy: Design and implementation of next-generation honeynet via SDN},
  author={Kyung, Sukwha and Han, Wonkyu and Tiwari, Naveen and Dixit, Vaibhav Hemant and Srinivas, Lakshmi and Zhao, Ziming and Doup{\'e}, Adam and Ahn, Gail-Joon},
  booktitle={2017 IEEE Conference on Communications and Network Security (CNS)},
  pages={1--9},
  year={2017},
  organization={IEEE}
}

@misc{gabrys2024honeyganpotsdeeplearning,
      title={HoneyGAN Pots: A Deep Learning Approach for Generating Honeypots}, 
      author={Ryan Gabrys and Daniel Silva and Mark Bilinski},
      year={2024},
      eprint={2407.07292},
      archivePrefix={arXiv},
      primaryClass={cs.CR},
      url={https://arxiv.org/abs/2407.07292}, 
}

@book{euler,
  title={A First Course in Rational Continuum Mechanics V1},
  author={Truesdell, Clifford A},
  year={1992},
  publisher={Academic Press}
}

@inproceedings{uitto2017survey,
  title={A survey on anti-honeypot and anti-introspection methods},
  author={Uitto, Joni and Rauti, Sampsa and Laur{\'e}n, Samuel and Lepp{\"a}nen, Ville},
  booktitle={World Conference on Information Systems and Technologies},
  pages={125--134},
  year={2017},
  organization={Springer}
}

@article{stefan-boltzmann,
  title={Hawking radiation, the Stefan--Boltzmann law, and unitarization},
  author={Giddings, Steven B},
  journal={Physics Letters B},
  volume={754},
  pages={39--42},
  year={2016},
  publisher={Elsevier}
}

@misc{raccoonos,
    title = {RACCOON OS},
    author = {José Manual Diez, Fabian Krech, Phillip Wüstenberg},
    howpublished = {\url{https://gitlab.com/rccn}},
    year = {},
    note = {[Accessed 20-01-2025]},
}

@misc{yamcs,
	author = {},
	title = {{Y}amcs {M}ission {C}ontrol},
	howpublished = {\url{https://yamcs.org}},
	year = {2025},
	note = {[Accessed 23-01-2025]}
}

@inproceedings{kaufeler1994esa,
  title={The ESA standard for telemetry and telecommand packet utilisation: PUS},
  author={Kaufeler, Jean-Francois},
  booktitle={NASA. Goddard Space Flight Center, Third International Symposium on Space Mission Operations and Ground Data Systems, Part 2},
  year={1994}
}

@misc{pusecosysytem,
    title = {CCSDS Mission Operations},
    author = {Sam Cooper, Mario Merri},
    howpublished = {\url{https://indico.esa.int/event/62/contributions/2797/attachments/2307/2667/1235_-_mission-operation-services---future-trends_Presentation.pdf}},
    note = {[Accessed 20-01-2025]}
}

@article{drmola2018kessler,
  title={Kessler syndrome: System dynamics model},
  author={Drmola, Jakub and Hubik, Tomas},
  journal={Space Policy},
  volume={44},
  pages={29--39},
  year={2018},
  publisher={Elsevier}
}

@incollection{boschetti2022space,
  title={Space cybersecurity lessons learned from the viasat cyberattack},
  author={Boschetti, Nicol{\`o} and Gordon, Nathaniel G and Falco, Gregory},
  booktitle={ASCEND 2022},
  pages={4380},
  year={2022}
}

@article{kavallieratos2023exploratory,
  title={An exploratory analysis of the last frontier: A systematic literature review of cybersecurity in space},
  author={Kavallieratos, Georgios and Katsikas, Sokratis},
  journal={International Journal of Critical Infrastructure Protection},
  pages={100640},
  year={2023},
  publisher={Elsevier}
}

@inproceedings{willbold2024vsaster,
  title={VSAsTer: Uncovering Inherent Security Issues in Current VSAT System Practices},
  author={Willbold, Johannes and Schloegel, Moritz and Bisping, Robin and Strohmeier, Martin and Holz, Thorsten and Lenders, Vincent},
  booktitle={Proceedings of the 17th ACM Conference on Security and Privacy in Wireless and Mobile Networks},
  pages={288--299},
  year={2024}
}
